\documentclass[usenatbib,useAMS]{mnras}

\usepackage[T1]{fontenc}
\usepackage{ae,aecompl}
\usepackage{graphicx}	
\usepackage{amsmath}	
\usepackage{amssymb}	
\usepackage{multicol}        
\usepackage{bm}		
\usepackage{pdflscape}	
\usepackage{natbib}
\usepackage{times}
\usepackage{float}
\usepackage{caption}
\usepackage{subcaption}
\usepackage{multirow}
\usepackage{color,soul}
\usepackage{booktabs,caption,fixltx2e}
\usepackage[flushleft]{threeparttable}

\newcommand \kpc        {\,{\rm kpc}}
\newcommand \sigmagas    {$\Sigma_{\rm \bld {gas }} $\ }
\newcommand \sigmatotalgas {$\Sigma_{\rm \bld {total\, gas }} $\ }
\newcommand \eqsigmagas    {\Sigma_{\rm \bld {gas }}}
\newcommand \sigmasfr     {$\Sigma_{\rm \bld {SFR }} $\ }
\newcommand \eqsigmasfr     {\Sigma_{\rm \bld {SFR }}}
\newcommand \sigmastar    {$\Sigma_{\rm \bld {star }} $\ }
\newcommand \eqsigmastar    {\Sigma_{\rm \bld {star }}}
\newcommand \halpha    {H$\alpha $\ }
\newcommand \um    {$\mu$m\ }

\newcommand \nprime {N$^\prime$}
\newcommand \eqnprime {N^\prime}
\newcommand \Spitzer {{\it Spitzer }}
\newcommand \Galex {GALEX }
\newcommand \Herschel {{\it Herschel }}

\begin{document}
\title[STAR FORMATION LAWS IN M31]{Star formation laws in the Andromeda galaxy: gas, stars, metals and the surface density of star formation}
\author[S. Rahmani, et. al.]{S.~Rahmani,\thanks{Contact e-mail: \href{mailto:srahma49@uwo.ca}{srahma49@uwo.ca}} S.~Lianou, P.~Barmby\\
Department of Physics $\&$ Astronomy, University of Western Ontario, London, ON N6A 3K7, Canada} 
\maketitle

\begin{abstract} 
 We use hierarchical Bayesian regression analysis to investigate star formation laws in the Andromeda galaxy (M31) in both local (30, 155, and 750~pc) and global cases. We study and compare the well-known Kennicutt-Schmidt law, the extended Schmidt law and the metallicity/star formation correlation. Using a combination of \halpha and 24 \um emission, a combination of far-ultraviolet and 24~$\mu$m, and the total infrared emission, we estimate the total star formation rate (SFR) in M31 to be between $0.35\pm 0.04$~M$_{\odot}$~yr$^{-1}$~and~$0.4\pm 0.04$~M$_{\odot}$~yr$^{-1}$. We produce a stellar mass surface density map using IRAC 3.6~$\mu$m emission and measured the total stellar mass to be $6.9 \times 10^{10}$~M$_{\odot}$. For the Kennicutt-Schmidt law in M31, we find the power-law index $N$ to be between 0.49 and 1.18; for all the laws, the power-law index varies more with changing gas tracer than with SFR tracer. The power-law index also changes with distance from the centre of the galaxy. We also applied the commonly-used ordinary least squares fitting method and showed that using different fitting methods leads to different power-law indices. There is a correlation between the surface density of SFR and the stellar mass surface density, which confirms that the Kennicutt-Schmidt law needs to be extended to consider the other physical properties of galaxies. We found a weak correlation between metallicity, the SFR and the stellar mass surface density.
\end{abstract}


\section{Introduction}
\label{sec:intro}

Understanding the underlying processes of formation of stars is a necessary step for explaining the formation and evolution of galaxies from the early universe to the current epoch, as well as the origin of planetary systems. Stars form when molecular cloud cores collapse; however, a complete picture explaining the physical processes involved in the formation of stars is yet to emerge. Various phenomena trigger the collapse of molecular clouds and star formation including the environment of the star-forming region, gas accretion and cooling, H$_2$ formation, the initial mass distribution and chemical compositions of the gas,  existence of dust, and other quantities in a galaxy. The physical processes leading to the formation of stars are naturally complex. Therefore, the basis for a theory of star formation requires a strong foundation of empirical data and/or observations.

The first attempt at scaling the star formation rate was to find a relation between the total star formation rate (SFR) and the mass of interstellar gas. \citet{Schmidt59} proposed a power-law relation between SFR and the gas mass ($M_{\rm gas}$) with {\it N} as the power law exponent. Since then, for more than 50 years empirical star formation laws (relations) have been investigated using both observational data and numerical simulations. Later, \citet{Kennicutt98a} examined H$\alpha$, H\,{\sc I}, and CO observations of 61 normal spiral galaxies and 36 starburst galaxies. He showed that the disk-average SFRs and gas densities for these samples are well represented by the Schmidt law with a power-law exponent of $1.4 \pm 0.15$. Using this power-law index, the Schmidt law can be rewritten as: 

\begin{equation}
\label{equ:ks_org}
\eqsigmasfr \propto \eqsigmagas^{1.4\pm0.15},
\end{equation}
\noindent which is often referred to as the Kennicutt-Schmidt (K-S) law, where \sigmasfr is the surface density of star formation, and \sigmagas is the surface density of gas in a galaxy. Moreover, \citet{Kennicutt07} investigated the K-S law in spatially resolved regions (0.5--2~kpc across) in the spiral galaxy M51. To calculate the SFR, they used a combination of 24~$\mu$m and H${\alpha}$ emission, and for the gas surface density, a constant conversion factor for CO to H$_2$ was applied. They showed that the K-S law also holds locally in this galaxy, and found {\it N} to be $1.56 \pm 0.04$.

The K-S law was applied to different types of both local and high-redshift galaxies \citep[e.g.][]{Boissier07,Kennicutt07, Bigiel08, Freundlich13}, but the results were never precise. Low gas density regions ($ < 1-10$~M$_{\odot}$~pc$^{-2}$), such as regions outside disks of galaxies, are an example of the regions for which the K-S law fails. The calculated SFR for these regions is much lower than the value predicted by the K-S law \citep[e.g.][]{Martin01, Bigiel08}. Using different instability models, \citet{Hunter98} showed that in low surface brightness galaxies, the current star formation activity correlates with stellar mass density. Furthermore, a correlation between the star formation rate and the stellar mass density was measured in many studies using both observational data and numerical simulations \citep[e.g.][]{Hunter04,Leroy08,Krumholz09,Shi11,Kim11,Kim13}. 
Studying a large sample of galaxies, \citet{Shi11} found a close relationship between stellar mass surface density, SFR and gas surface density.  They showed this relation as a power law and referred to it as the extended Schmidt law:
\footnote{\citet{Shi11} introduced the extended Schmidt law as a power-law relation between the star formation efficiency (surface density of SFR per surface density of gas, SFE) and $\eqsigmastar$, but to be consistent with equation~\ref{equ:ks_org}, we are using equation~\ref{equ:es_org}, which is equation 5 in their paper.} 

\begin{equation}
\label{equ:es_org}
\eqsigmasfr \propto \eqsigmagas^{\eqnprime} \eqsigmastar^{\beta},
\end{equation}
\noindent where \sigmasfr is the SFR surface density while \sigmagas and \sigmastar are the gas mass and stellar mass surface densities,  $\eqnprime$ and $\beta$ are the indices found by \citet{Shi11} to be $1.13 \pm 0.05$ and $0.36\pm0.04$, respectively, in the global case. By testing this relation in sub-kiloparsec resolution regions in 12 spiral galaxies, they showed that the extended Schmidt law works just as well as the K-S law in spatially resolved regions ($\sim$1 kpc). Furthermore, they found the power-law indices for the local regions to be $\eqnprime = 0.80 \pm 0.01$ and $\beta = 0.63\pm0.01$, and concluded that this law is acceptable for low surface brightness galaxies and regions where the K-S law fails.

A galaxy's metallicity is another physical quantity that likely plays a role in the rate of star formation. However, it affects each component of star formation laws in a different way, making it complicated to fully understand its implications. There are many studies on correlation of metallicity with stellar mass and SFR, as well as the effect of metallicity on the ISM \citep[e.g.][]{Boissier03, Leroy08, Krumholz09, Mannucci10, Dib11a, Lilly13}. These studies suggest that metallicity has a critical role in the formation of H$_2$ gas. Thus there should be a correlation between metallicity and the SFR. \citet{Krumholz09} introduced a theoretical SFR law which suggests that the SFR and metallicity of galaxies have a power-law correlation, but the power-law index changes with the amount of total gas. 
 Using results obtained from both observations and analytical models, \citet{Mannucci10} and \citet{Lilly13} showed that metallicity depends on both stellar mass and the SFR, which they referred to as a 'fundamental metallicity relation'. They found that at any fixed stellar mass, metallicity and the SFR have an inverse correlation. It should be noted that the above two correlations are only valid in the global case. However, \citet{Leroy08} showed that the SFR as measured by combining far-UV and 24$\mu$m emission does not change with changing metallicity. \citet{Roychowdhury15} argued that in nearby spiral galaxies effect of metallicity on the SFR calculation is smaller than the calibration errors and can be ignored. On the other side, \citet{Boissier03} showed that metallicity has an inverse correlation with CO-H$_2$ conversion factor. Therefore, using a constant conversion factor at high metallicity regions causes the overestimation of the molecular gas mass, and in low metallicity it leads to an underestimation.

In order to investigate star formation laws, the first step is to calculate the SFR. Many studies are devoted to determining how flux measurements of star-forming regions can be accurately translated into the rate of formation of new stars \citep[e.g.][]{Kennicutt12, Calzetti13, Zhu08, Kennicutt09, Boquien10, Boquien11, Hao11}. \citet{Kennicutt98b} calibrated luminosity of galaxies as a way of measuring the SFR in specific wavelengths using relations between the SFR per unit mass or per unit luminosity and the integrated colour of the system provided by synthesis models \citep[e.g.][]{Bruzual93}. Subsequent studies tried to find a similar calibration for measuring the SFR of galaxies in other bands. \citet{Kennicutt09} and \citet{Hao11} introduced new SFR calibrations using a combination of \halpha or far-ultraviolet (FUV) and far-infrared (FIR) emission. In addition to using suitable calibrations for each region, considering the differences between the global and the local ($\sim 0.5-1$~\kpc) cases is important. In $\S$~\ref{sec:sfr}, we will discuss the SFR in the Andromeda galaxy and how it was determined. 

Calculating the surface density of gas is the next step in testing star formation laws.
Considering the effect of the interstellar medium (ISM) is essential in understanding star formation laws since stars are born from gas and also release their material into the ISM when they reach the end of their evolution. 
A map of the total gas in the ISM is produced by direct observations of the gas or by using interstellar dust as a tracer. Neutral and molecular hydrogen are the most common elements in the ISM. Therefore, for producing the map of the total gas in galaxies, maps of these two components can be added together and multiplied by a constant factor to account for heavier elements. 
Calculations of the gas surface density for the Andromeda galaxy are shown in $\S$~\ref{sec:ISM}. 

Additionally, for testing the Extended Schmidt law, measuring the stellar mass density is necessary. Measuring the mass of stellar populations is indirect and subject to significant uncertainties. One method to calculate the stellar mass within a galaxy is based on stellar population models \citep[e.g.][]{ Bruzual93, Kotulla09}. These models are used to connect stellar mass to an observable quantity such as luminosity in different wavelengths, colours, spectral energy distribution from spectroscopy or multi-band observations.
Having various methods to measure stellar mass helps to compare the results and determine whether there are any systematic differences. 
\citep{McLaughlin05} found differences between methods to be on the order of a few tens of percent and concluded that one should be careful to model subtleties.
Section $\S$~\ref{sec:starmass} describes our calculations of stellar mass in the Andromeda galaxy.

In this paper, we present our results from testing and comparing both the K-S law and the Extended Schmidt law on the Andromeda galaxy (M31). Additionally, we applied these two laws on three different regions in M31 to determine whether there is any dependence on distance from the centre of the galaxy. Since M31 is the nearest spiral galaxy to our own, high resolution images of this galaxy are available, providing data from various regions with different physical properties (e.g. metallicity, surface brightness, gas density). This inside look helps us test star formation laws in diverse physical conditions. Thus M31 is a suitable testbed for studying scaling laws.

Furthermore, the range of power-law indices calculated for the Andromeda galaxy using the K-S law is between 0.5 and 2 \citep[e.g.][]{Tabatabaei10,Ford13}. The use of different methods and data results in measuring different values for the power-law index. In a recent study of the K-S law in M31, \citet{Ford13} tested this law on six annular regions using three different ISM maps (H$_2$ only, total gas calculated from H$_2$ plus H\,{\sc I}, and total gas calculated from dust emission). The measured power-law indices for each map and region vary between 0.6 to 2.03. The origin of these variations in the results is still an open question. It is unclear whether it depends on galactocentric distance, metallicity, gas tracers, SFR tracers, fitting methods, or because of the K-S law not working in M31. In order to examine the dependence of the results on the fitting method, we also applied a new statistical method, instead of the Ordinary Least Squares (OLS) fitting which is more commonly used in the literature to test the K-S law ($\S$~\ref{sec:fitting}).

\section{Data}
\label{sec:data}
Being the closest large galaxy to the Milky Way, M31 has extensively been studied using both ground- and space-based telescopes. Therefore, large datasets on M31 exist spanning from gamma-ray to radio wavelengths, allowing us to measure all the required parameters to test star formation laws in this galaxy. Table~\ref{table:data} lists the data we used in this paper. Each dataset comes with a different angular resolution and pixel size. In order to match the images at different wavelengths, we smoothed the maps to the same FWHM using the kernel library and convolution code of \citet{Aniano11}. Our final results have two different angular resolutions depending on which gas tracer we use in studying the SFR laws. For studying the correlations between the SFR and molecular gas, our maps have the same resolution and pixel size as the $^{12}$CO(J:$1\rightarrow0$) data, while for investigating the relationships between the SFR and atomic gas (or total gas), we smoothed and re-gridded our maps to the resolution and pixel size of the atomic gas map.

\begin{table*}
\centering
\caption{Data used in this study.}
\label{table:data}
\begin{tabular}{@{}lcccc}
\hline\hline
Wavelength & FWHM & Coverage area &Telescope
& Ref. \\
\hline
1550~\AA & 4\arcsec.5 & 5\degr $\times$ 5\degr &GALEX & \citet{Martin05}\\  
2250~\AA & 6\arcsec & 5\degr $\times$ 5\degr &GALEX & \citet{Martin05}\\
6570~\AA  & 1\arcsec & 2\degr.2 $\times$ 0\degr.6 &KPNO& \citet{Massey07}\\
3.6~\um & 1\arcsec.7 & 3\degr.7 $\times$ 1\degr.6 &\Spitzer & \citet{Barmby06} \\ 
8~\um & 1\arcsec.9 & 3\degr $\times$ 1\degr &\Spitzer & \citet{Barmby06} \\ 
24~\um & 6\arcsec & 3\degr $\times$ 1\degr &\Spitzer & \citet{Gordon06} \\ 
70~\um & 18\arcsec & 3\degr $\times$ 1\degr &\Spitzer & \citet{Gordon06} \\
160~\um & 12\arcsec & 5\degr.5 $\times$ 2\degr.5 &\Herschel & \citet{Fritz12} \\
350~\um & 24\arcsec.9 & 3\degr $\times$ 1\degr &\Herschel & \citet{Groves12} \\
2.6~mm & 23\arcsec & 2\degr $\times$ 0\degr.5 &IRAM 30-m & \citet{Nieten06}\\
21~cm & 60\arcsec $\times$ 90\arcsec & 5\degr.2 $\times$ 1\degr.5 &DRAO & \citet{Chemin09}\\
\hline
\end{tabular}
\end{table*}

Several different observational datasets are used to compute properties of M31 for testing star formation laws.
One method for calculating the SFR uses FUV emission as a tracer of the star-forming regions: for this we
use the FUV image of M31 as observed by the Galaxy Evolution Explorer \citep[GALEX;][]{Martin05}. 
A second method to calculate the SFR involved using a combination of \halpha and 24~$\mu$m emission. \citet{Massey06, Massey07} mapped M31 in broad and narrow bands, including \halpha, as part of the Nearby Galaxies Survey using the KPNO 4-m telescope. A detailed description of how our \halpha map was made can be found in Appendix~\ref{app:halpha}.

Calculations of the SFR using the total infrared emission requires observations at 8, 24, 70, and 160~\um, made
with the {\em Spitzer} \citep{Werner04} and {\em Herschel} \citep{Pilbratt10}  space telescopes. 
Observations with the Infrared Array Camera (IRAC; \citep{Fazio04}) are made in four channels (3.6, 4.5, 5.8, and 8~\um); IRAC observations of M31 were obtained by \citet{Barmby06}, covering a region $3\degr.7 \times 1\degr.6$. 
The Multiband Imaging Photometer for \Spitzer (MIPS) observed M31 at 24, 70, and 160 \um and covered a region $\sim 3\degr \times 1\degr$ \citep{Gordon06}. {\em Herschel} observations of M31 were made with with PACS \citep[Photodetector Array Camera and Spectrometer;][]{Poglitsch10}  at 100 and 160~\um and SPIRE  \citep[Spectral and Photometric Imaging Receiver;][]{Griffin10} 
at 250, 350, and 500~\um, covering a region  $\sim 5\degr.5 \times 2\degr.5$ \citep{Fritz12}.
The IRAC 8, MIPS 24, MIPS 70, and PACS 160 images were used to calculate M31's total infrared emission. 
The IRAC 3.6~\um band image was used to calculate the stellar mass. 

Gas densities in galaxies are derived from observations at millimetre and centimetre radio wavelengths.
Emission from the $^{12}$CO (J:$1\rightarrow0$) line observed on-the-fly with the IRAM 30-m telescope \citep{Nieten06} was used to calculate the H$_2$ column density of M31's interstellar medium. This map covers a $2\degr \times 0\degr.5$ region with an angular resolution of $23\arcsec$ and has the smallest coverage of the galaxy among the maps we used in this paper. In this study, we also used a 21-cm emission map of the atomic gas (H\,{\sc I}) in M31 from \citet{Chemin09}. This map was made using the synthesis telescope at the Dominion Radio Astrophysical Observatory and covers a $5\degr.1 \times 1\degr.5$ region with an angular resolution of $60\arcsec \times 90\arcsec$.

\section{Measuring the components of Star Formation Laws}
Before studying the star formation laws, we need to calculate the components of these laws. We used different data and calibrations to produce the SFR maps, the stellar mass surface density and the gas mass surface density. In this section we describe how we measured each ingredient using available data, and calibrations. To use these calibrations, we assumed the M31 distance to be 0.78~Mpc~\citep{McConnachie05}. The ``total'' values recorded in this paper are integrated out to a distance of~25~kpc from the centre of the galaxy along the major axis. 

\subsection{Stellar Initial Mass Function}
\label{sec: imf}
In order to determine star formation rate and stellar mass from calibrated photometric data, it is necessary to know the IMF. 
Stellar population synthesis models make different assumptions about the IMF, often either \citet{Salpeter55},
$ dN / d \log M \propto M^{-3.5 }$ over the full range of masses, or 
\citet{Kroupa01} $ dN / d \log M \propto M^{-\Gamma }$ 
with $\Gamma=-2.3$ at low masses and  $\Gamma=-3.3$ for higher mass.
For the same total mass, the Salpeter and Kroupa IMFs produce almost the same amount of high-mass stars, but the Salpeter IMF produces more low-mass stars. Different predictions of the number of stars between various IMFs are one of the main uncertainty sources in the calculation of the stellar mass \citep{Eskew12,Brewer12}. \citet{Eskew12} showed that the total mass of low-mass stars can change by a factor of 2.5 depending on the IMF.
\citet{Calzetti13} showed the impact of different IMF assumptions on SFR calibrations. Adopting a modified \citet{Kroupa01} IMF, with its maximum stellar mass set to 30 M$_{\odot}$ instead of 100 M$_{\odot}$, changes the SFR calibration constant by factors 1.4 to 5.6 for different SFR indicators. Changes in the IMF mostly affect the calibration based on \halpha luminosity, because a significant amount of the ionizing photons comes from stars more massive than $\sim 20$~M$\sun$.

In this project we used values based on the Kroupa IMF as is standard in the field, although discussion continues on whether the IMF is universal \citep{Bastin10}. Where necessary we converted from calibration values based on the Salpeter IMF by multiplying by 0.67  \citep{Madau14}.

\subsection{Star Formation Rate}
\label{sec:sfr}
\subsubsection{FUV plus 24~$\mu$m Star Formation Rate}

Our first star formation map was produced using a calibration of a combination of FUV emission and 24 \um emission introduced by \citet{Hao11}. Since the peak of the emission of massive and young stars (O-B and A type) is in the UV part of the spectrum, this wavelength is one of the most commonly used star formation rate indicators \citep[e.g.][]{Kennicutt89}. The UV emission traces recently formed stars ($\sim 100$ Myr) \citep[e.g.][]{Kennicutt98a, Calzetti05}. However, the downside of only using the FUV light as a tracer of the star-forming region is that this emission is very sensitive to dust extinction. On the other hand, 24 \um emission is dominated by emission from dust heated by UV photons from young and hot stars. This emission is sensitive to the star formation timescale of $\la10$ Myr \citep{Calzetti07}. The FUV band of \Galex and the 24 \um band of \Spitzer were used to measure the SFR for stars with ages $<100$~Myr calibrated by \citet{Hao11}  using nearby galaxies and the \citet{Kroupa03} IMF as:
\begin{equation}
\label{equ: fuvplus24}
SFR =4.46\times10^{-44}[L(FUV)_{\rm obs}+3.89L(24\mu m)],
\end{equation}
\noindent where SFR is in M$_{\odot}$~yr$^{-1}$, and L(24$\mu$m) and L(FUV)$_{\rm obs}$ are in erg~s$^{-1}$. L(FUV)$_{\rm obs}$ is the luminosity of the galaxy in the FUV. The subscript "obs" indicates that the FUV emission is not corrected for extinction. Nevertheless, this emission was corrected for the foreground stars' emission using a method introduced by \citet{Leroy08}. We masked out pixel values in the FUV$_{obs}$ map for which I$_{NUV}$/I$_{FUV}$ $>$ 15, assuming that those pixels are dominated by foreground emission. This process results in masking about 3\% of the pixels within a 25 kpc radius from the centre in the FUV$_{obs}$ map, which leads to a slight underestimate of the total SFR. 
Then we smoothed the FUV$_{obs}$  map to have the same angular resolution and pixel size as the MIPS 24~$\mu$m, and masked the same regions in the 24 \um map as in the FUV$_{obs}$  map. We added all the pixels in the map and calculated the total SFR to be $0.31\pm 0.04$~M$_{\odot}$~yr$^{-1}$. The reported uncertainty for the total SFR only considers the values we measured and does not include uncertainties related to the masked pixels.

\begin{figure*}
    \centering
       \includegraphics[width=164mm]{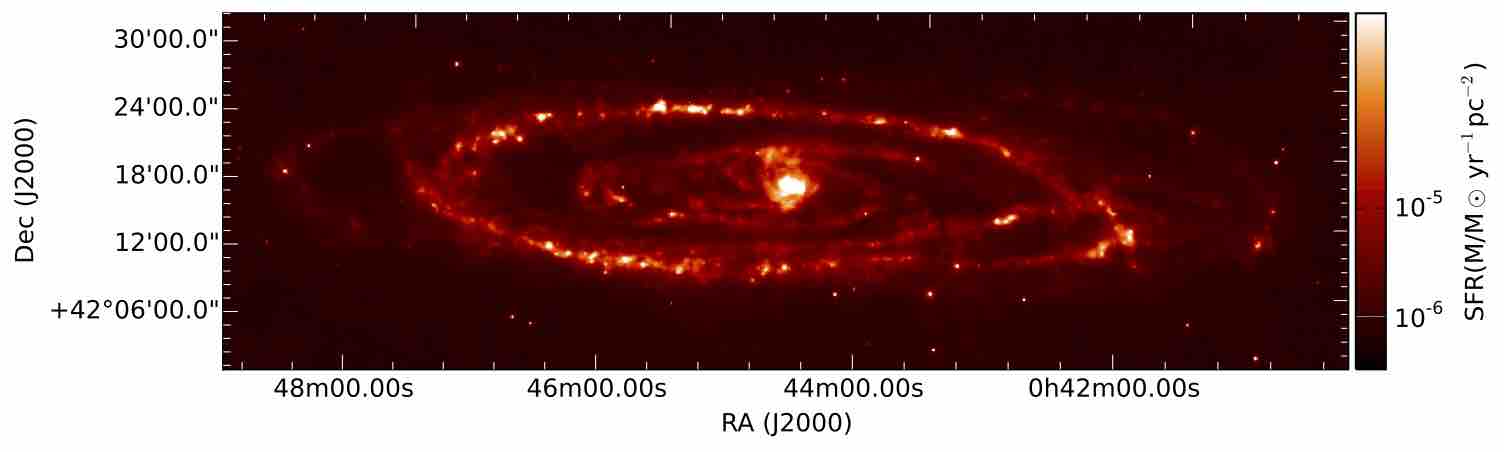}
    \caption {SFR map calculated from total infrared emission (see Section~\ref{sec:sfr_fir}). The SFR map here has an angular resolution of 18\arcsec and pixel size of 9.85\arcsec, the same as the MIPS 70 $\mu$m map.}
    \label{fig:sfrs}
\end{figure*}

\subsubsection{\halpha plus 24 \um Star Formation Rate}
\label{sec:sfr_halpha}

The SFR can also be calculated using a combination of \halpha and \Spitzer  24 \um maps. \halpha emission of galaxies is dominated by light emitted from young and massive stars. Star formation timescales traced by this emission are $\sim 6-10$ Myr \citep[e.g.][]{Kennicutt09, Calzetti13}. Therefore, \halpha alone could be used to calibrate the SFR \citep[e.g.][]{Osterbrock06, Kennicutt09}. However, \halpha emission is very sensitive to dust extinction. Therefore, the combination of the \halpha and 24\um of considered a better choice for calculating the SFR. We used a calibration introduced by \citet{Kennicutt09} to measure the SFR:
\begin{equation}
\label{equ: halphaplus24_g}
SFR = 5.5 \times 10^{-42}[L(H{\alpha})_{\rm obs} + 0.020L(24\mu m)],
\end{equation}
\noindent where L(H${\alpha}$)$_{\rm obs}$ is the observed \halpha luminosity without correction for internal dust attenuation, given in the unit of erg~s$^{-1}$. L(24$\mu$m) is the $24\mu$m IR luminosity in erg~s$^{-1}$, and SFR is in M$_{\odot}~$yr$^{-1}$. It was indicated that the formulation above can only be used for the global situation, i.e., the total SFR.

\citet{Calzetti07} introduced a new calibration using the same bandpasses for the case of local regions:
\begin{equation}
\label{equ: halphaplus24_l}
SFR = 5.5 \times 10^{-42}[L(H{\alpha})_{obs} + 0.033L(24\mu m)],
\end{equation}
\noindent where the units here are the same as in equation~\ref{equ: halphaplus24_g}. Equation~\ref{equ: halphaplus24_l} is useful for regions less than $\sim 1$ kpc and is what we used for testing the SFR laws in local regions. The total M31 SFR calculated from summing pixel values and applying equation~\ref{equ: halphaplus24_g} is $0.35 \pm 0.01$~M$\sun$~yr$^{-1}$, which is in good agreement with previous studies.

\subsubsection{Total Infrared Star Formation Rate}
\label{sec:sfr_fir}
The total infrared emission of a system can also be used as a tracer of the SFR. Dust absorbs radiation from hot and young stars and re-emits it at infrared wavelengths; however, there is no one-to-one mapping between IR and UV emission. Therefore, integrating over the full wavelength range of the IR part and calculating the TIR luminosity is a better tracer of the SFR compared to IR single-band emission \citep{Calzetti13}. \citet{Calzetti13} assumed that stellar bolometric emission is completely absorbed and re-emitted by dust, i.e., $L_{star}(bol) = L(TIR)$, and calibrated the TIR luminosity of a system to calculate the SFR for a stellar population undergoing constant star formation over 100 Myr:
\begin{equation}
\label{equ:sfr_fir}
SFR(\rm TIR) = 2.8 \times10^{-44}L(\rm TIR),
\end{equation}
\noindent where SFR(TIR) and L(TIR), the star formation rate calculated from TIR emission and TIR luminosity, are in M$_{\odot}$~yr$^{-1}$ and erg~s$^{-1}$, respectively. Part of the dust emission in galaxies (specially in M31) comes from dust heated by the cosmic background radiation \citep[e.g.][]{Dole06, Calzetti13, Mattsson14}, hence, using equation~\ref{equ:sfr_fir} slightly overestimates the SFR in M31.  

Total infrared luminosity can be measured from the integration of the infrared part of a galaxy's spectral energy distribution or by combining photometric data at different IR wavelengths. As an example of the second approach, \citet{Draine07} modelled \Spitzer data and calibrated IR single-band photometry data to calculate the TIR luminosity. \citet{Boquien10} tested and modified this calibration as:
\begin{equation}
 \label{equ: TIR}
L(\rm TIR) = 0.95L(8) + 1.15L(24) + L(70) + L(160),
\end{equation}
\noindent where L(8), L(24), L(70), and L(160) are the luminosities of the galaxy at 8 $\mu$m, 24~$\mu$m, 70~$\mu$m, and 160~\um in units of erg~s$^{-1}$. We used results from equation~\ref{equ: TIR} in the equation~\ref{equ:sfr_fir} to calculate our final map of the M31 SFR (Figure~\ref{fig:sfrs}). Using this method, we calculated the total SFR by adding all the pixels to be $0.40 \pm 0.04$~M$_{\odot}$~yr$^{-1}$. Table~\ref{table:sfr} compares recent measurements of the total SFR of M31 from the literature and the present work.

\begin{table*}
\begin{minipage}{100mm}
\caption{Comparison of the total star formation rate of M31}
\label{table:sfr}
\begin{tabular}{@{}lcc}
\hline\hline
Ref.&Method&SFR(M$_{\odot}$yr$^{-1}$) \\
\hline
Current work&FUV and 24~$\mu$m&0.31 \\
Current work&\halpha and 24~$\mu$m&0.35 \\
Current work&TIR luminosity&0.4\\
\citet{Ford13}&FUV and 24~$\mu$m&0.25\\
\citet{Ford13}&TIR luminosity&0.48-0.52\\
\citet{Azimlu11}& \halpha and 24~$\mu$m&0.34\\
\citet{Azimlu11}&Extinction-corrected \halpha&0.44\\
\citet{Tabatabaei10}&Extinction-corrected \halpha&0.27--0.38\\
\citet{Barmby06}&Infrared 8\um luminosity& 0.4\\
\hline
\end{tabular}
\end{minipage}
\end{table*}


\subsection{Gas Surface Density}
\label{sec:ISM}

The molecular form of hydrogen is very difficult to detect. Therefore, CO (usually the $J(1\rightarrow 0)$ rotational transition, observed at 2.6 mm) is used as a tracer of the mass of the molecular cloud dominated by molecular hydrogen \citep[see, for example,][]{Sanders84}. Higher rotational transitions of CO can be used as tracers as well; however, they are not as common as $J(1\rightarrow 0)$. Equation~\ref{equ:conversion} shows the relation between CO emission and H$_2$ column density:
\begin{equation}
\label{equ:conversion}
\rm N_{H_2}/\rm cm^{-2} = X_{CO} \times I_{CO}/[{\rm K km s^{-1}}],
\end{equation}
\noindent where X$_{CO}$ is the conversion factor (also known as the X-factor), I$_{CO}$ is the CO intensity in  K~km~s$^{-1}$ and N$_{H_2}$ is the molecular hydrogen column density. The X-factor could be different in regions within a galaxy due to varying metallicity \citep{Wilson95, Bosselli02, Bolato13}. In the case of M31, different values of X-factor in the range of ($1-5.6 \times 10^{20}$) were adopted in previous studies \citep[e.g.][]{Ford13, Bolato13, Leroy11, Bolato08, Nieten06}; however, since any constant difference in X-factor leads to horizontal changes in a plot of log(SFR) versus log($\Sigma_{\rm {gas}}$) and does not affect our power-law indices for molecular gas, we chose $X_{\rm {CO}}= 2 \times 10^{20}$. This is the same value that is adopted in many other works \citep[e.g.][]{Ford13, Smith12}. Since H\,{\sc I} mass dominates over the H$_2$ mass in M31, the constant X-factor has a negligible effect on the total gas mass.

To calculate the mass of H$_2$ from the molecular hydrogen column density, we estimated the volume and volume density of H$_2$ in each pixel. Then, using M$_{H_2} = \mu_{H_2}m_H\rho_{H_2}$ where $ \mu_{H_2}$ is the mean molecular mass of H$_2$, m$_H$ is the mass of one hydrogen atom and $\rho_{H_2}$ is the volume density, we calculated M$_{H_2}$. 

The total gas mass was calculated from:
\begin{equation}
\label{equ:total_gas}
\rm M_{\rm \bld {total \, gas}} = 1.36[M_{\rm H I}+M_{H_2}],
\end{equation}
\noindent where the factor of 1.36 is a constant that takes into account the contribution of He and other heavier elements to the total gas mass, M$_{\rm H\,{\sc I}}$ is the mass of neutral hydrogen that was obtained from 21 cm observations (the map, which we got from \citet{Chemin09}, has units of number density which we used to calculate mass density of H\,{\sc I}), and M$_{H_2}$ was calculated from equation~\ref{equ:conversion} in units of M$_{\sun}$/pc$^2$. We assumed a gas mass of zero for regions where data for  either M$_{\rm H\,{\sc I}}$ or M$_{H_2}$ was not available, due to differences in coverage or bad pixels, etc. In this case the above equation gives a lower limit for the total gas; this applies to regions beyond 18~kpc from the galaxy centre.

\subsection{Stellar Mass}
\label{sec:starmass}
\begin{figure*}
\centering
\includegraphics[width=164mm]{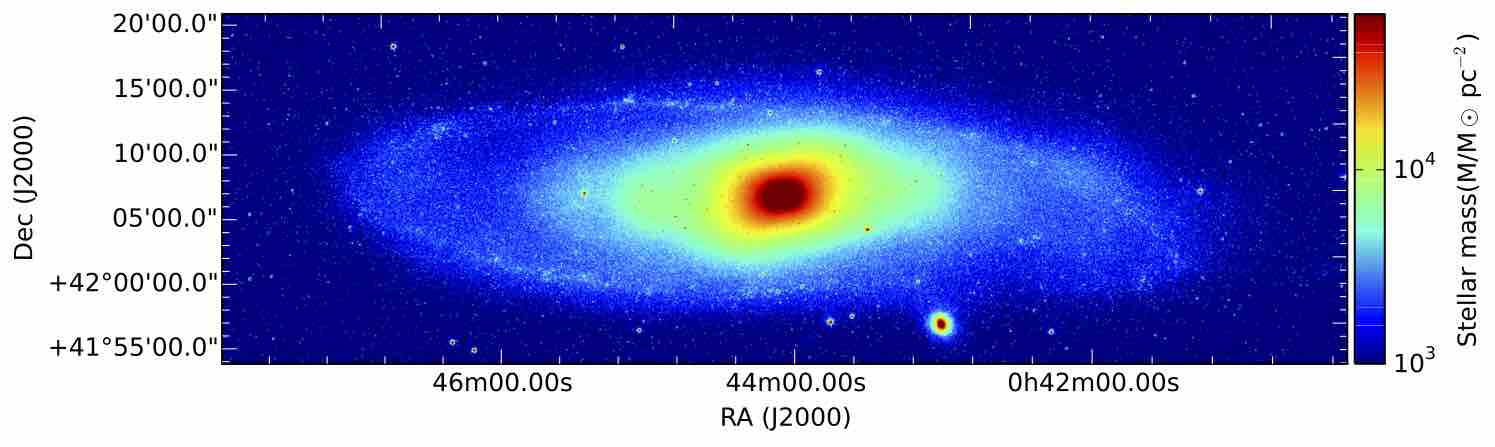}.
\caption{The stellar mass surface density map produced using IRAC 3.6~$\mu$m data and the calibration presented in equation~\ref{equ:eskew}.}
\label{fig:stellarmass}
\end{figure*}

Emission in near-infrared bands, such as 3.6~$\mu$m, 4.5~\um and $K$-band, is almost insensitive to emission from young stellar populations, dust absorption and emission, which makes these bands good tracers of the mass of older stellar populations \citep[e.g.][]{Elmgreen84, Eskew12, Zhu10}.We calculated the stellar mass within M31 using a calibration of the 3.6 \um flux by \citet{Eskew12}. They used data from the Large Magellanic Cloud and introduced an empirical relation between stellar mass and flux:
\begin{equation}
\label{equ:eskew}
M _{\star}= 10^{5.97} F_{3.6}\left(\frac{D}{0.05}\right)^2,
\end{equation}
\noindent where M$_{\star}$ is the stellar mass in solar masses, $F_{3.6}$ is the 3.6~\um flux in Jy, and $D$ is the distance to the galaxy in Mpc. This calibration is based on the \citet{Salpeter55} IMF. Since the Kroupa IMF is adopted for the SFR maps, we corrected the calibration to the Kroupa IMF (see Section~\ref{sec: imf}). We should note that \citet{Eskew12} gave two calibrations for calculating the stellar mass in their paper: one is equation~\ref{equ:eskew} which only uses 3.6~\um emission and the other uses both 3.6~\um and 4.5~\um emission. We noticed that the second calibration can only be used in the case of total stellar mass due to the fact that the correlation shows a non-linear dependency of stellar mass on $F_{3.6 \mu m}$ and $F_{4.5 \mu m}$, though this was not mentioned in the aforementioned paper. 
Figure~\ref{fig:stellarmass} shows the stellar mass map for M31 using IRAC 3.6~\um in the original angular resolution and pixel size of the IRAC 3.6~\um image. Using the coverage map of the IRAC 3.6~\um data, we used only pixels with coverage of more than 2 images per sky position, and replaced the rest with zero M$_{\odot}$. The replaced pixels include less than 8\% of the data;  since these pixels are mainly  in regions with low surface brightness, their effect on the final total mass is less than 1\%.
Using this method we calculated the total stellar mass of the galaxy by adding all pixel values and found it to be $6.9 \times 10^{10}$M$_{\odot}$ $\pm$ 6\%. This result is in fairly good agreement with the results of \citet{Tamm12}. They calculate the stellar mass of M31 to be $(10-15) \times 10^{10}$M$_{\odot}$, 56\% of which is in the disk with the rest in the bulge of the galaxy. In their recent work on the stellar mass of M31, \cite{Viaene14} measured the total stellar mass using the integrated fluxes to be $5.49 \times 10^{10}$M$_{\odot}$, which is in good agreement with our result.

\subsection{Metallicity}
\label{sec:metal}
 
The metallicity of a galaxy can be estimated by measuring abundances of heavier elements in its ISM. 
It is common to use  gas phase oxygen abundance, $[{\rm O/H}]$, as a gauge to determine metallicity \citep[e.g.][]{McGaugh91, Zaritsky94}. 
Results from modelling dust in the Milky Way and nearby galaxies showed that the dust to gas mass ratio depends on $[{\rm O/H}]$ and, consequently, on metallicity \citet{Draine07}. \citet{Draine14} assumed that depletion of gas elements from the gas to the diffuse ISM in M31 is similar to the Milky Way, and proposed that: 
 \begin{equation}
 \label{equ: metal}
\frac{M_d}{M_H}=0.0091 \frac{Z}{Z_{\odot}},
 \end{equation} 
\noindent where M$_d$ and M$_H$ are the total masses of dust and hydrogen, $Z$ is the metallicity of M31 and $Z_{\odot}$ is solar metallicity. 

\citet{Draine14} produced two dust maps of M31 using MIPS-160 resolution (FWHM  39\arcsec; 18\arcsec $\times$ 18\arcsec pixels) and SPIRE 350 resolution (FWHM 24\arcsec.9; 10\arcsec $\times$ 10\arcsec pixels), which are publicly available. 
We used the dust map with SPIRE 350 resolution, convolved and re-gridded  to have the same resolution and pixel size for the H\,{\sc I} map, then divided this dust mass by the mass of hydrogen in M31. 
However, in a private correspondence with the author, Draine pointed out that based on the Planck observations of Milky Way dust \citet{Tauber10}, the dust mass surface density map estimated for M31 by \citet{Draine14} may be systematically high by a factor of 2. We divided our dust map by 2 to account for this correction. 
We then generated a metallicity map of M31 by using the dust map in combination with equation~\ref{equ: metal}. \citet{Draine14} showed that with increasing distance from the centre of M31, the dust to H mass ratio declines monotonically and concluded that metallicity also behaves in the same way. 
Their fitting results showed that M31 can be divided into three regions ($R< 8~\kpc$, $8~\kpc < R < 18~\kpc$, and $18~\kpc < R \la 25~\kpc$) where the dust to H mass ratio (or metallicity) has a different correlation with galactocentric distance.
  
\section{Scaling the star formation rate}
\subsection{Fitting method}
\label{sec:fitting}

\begin{figure*}
\centering
\includegraphics[width=164mm, height=218mm]{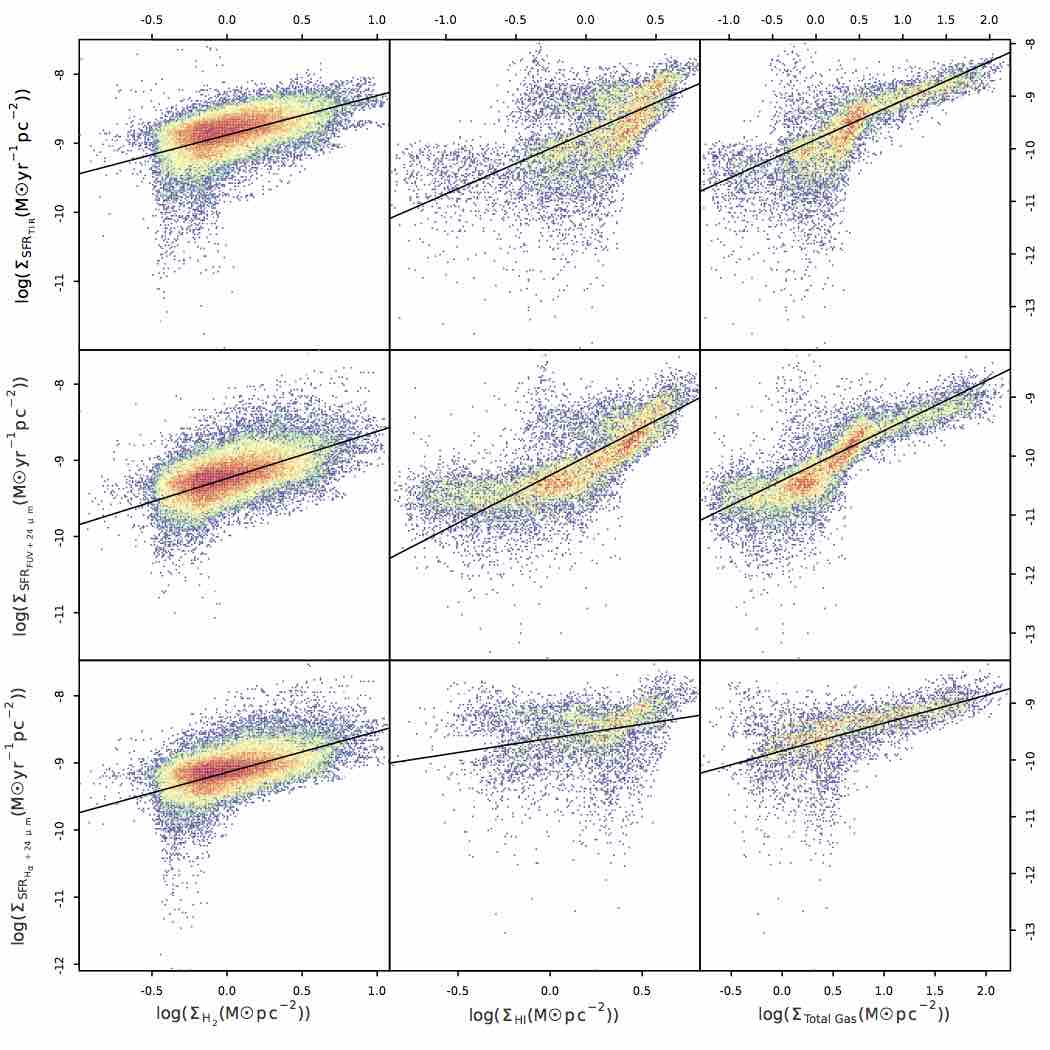}
\caption{Results from fitting the Kennicutt-Schmidt law to data from the whole galaxy using the pixel-by-pixel method. The colour represents the density of the pixels, increasing from blue to red. The number of points among the three columns is different due to the different angular resolution of the H\,{\sc I} and H$_2$ maps. Each point on the plots where the surface density of H$_2$ is used as a tracer of gas mass represents a region of size $\sim$30~pc (for maps with resolution of 23\arcsec) while each point for which the surface density of H\,{\sc I} or total gas mass is a tracer represents a region of size $\sim$155~pc (for maps with resolution of $60\arcsec \times 90\arcsec$). Solid lines show the best fit, using the mean value of the ranges in Table~\ref{table:res}.}
\label{fig:ks_all}
\end{figure*}

Producing the SFR, gas mass, stellar mass, and metallicity maps provides enough data to examine and compare the K-S law and the extended Schmidt law. To create surface density maps, we divided the value of each pixel by its area in pc$^2$. We investigated the K-S law and the extended Schmidt law using the pixel-by-pixel method in two ways. In one approach we used all the available pixels over the galaxy; in  the other approach we compare the laws in three elliptical regions at different distances from the centre of the galaxy as described in Section~\ref{sec:metal}, assuming a position angle of 38\degr and inclination of 77\degr\ for M31.
As mentioned in Section~\ref{sec:data}, since the H$_2$ gas mass map has higher angular resolution and smaller pixel size than the H\,{\sc I} gas mass map, our final results have two sets of different  angular resolutions and pixel sizes. 

Equations~\ref{equ:ks_org}~and~\ref{equ:es_org} can be re-written in logarithmic form to obtain two linear equations:
\begin{subequations}
\begin{equation}
\label{eq:sfr_law_ks_log}
\log_{10} \eqsigmasfr = N~\log_{10} \eqsigmagas + A,
\end{equation}
\begin{equation}
\label{eq:sfr_law_es_log}
\log_{10} \eqsigmasfr = \eqnprime~\log_{10} \eqsigmagas + \beta~\log_{10}\eqsigmastar  + A^\prime,
\end{equation}
\end{subequations}
\noindent where, {\it N}, $\eqnprime$, $\beta$, A, and A$^\prime$ are free fitting parameters. The units in the above equations are the same as those in equations~\ref{equ:ks_org} and ~\ref{equ:es_org}.

We found the free parameters by applying the hierarchical Bayesian linear regression method as described in \citet{Shetty13}. Shetty and colleagues used a Bayesian linear regression approach in a new method to find the K-S law parameters, considering the measurement uncertainties as well as hierarchical data structure.
We used the same technique as \citet{Shetty13} to estimate the K-S law parameters. For the extended Schmidt law, we extended the code such that instead of using  simple hierarchical Bayesian linear regression, which is used for the K-S law, it uses multiple hierarchical Bayesian linear regression. In this case, we were able to examine the effect of stellar mass on the SFR as shown in equation~\ref{eq:sfr_law_es_log}. 

Uncertainties in the SFR and stellar mass maps  were measured following the method described in \citet{Kennicutt07}. We used the quadratic sum of the variance of the local background for each luminosity map from the original pixel size images, Poisson noise of images, and calibration uncertainties. Note that this method measures the lower limit of the uncertainties for these variables. Uncertainties in H$_2$ mass were measured using the mean rms values of the  CO intensity, reported in \cite{Nieten06}. Uncertainties of the total gas map are calculated as the quadrature sum of the uncertainties in  H$_2$ and H\,{\sc I} masses. 

We also tested each law on three different regions which were chosen in the same manner as explained in Section~\ref{sec:metal}. Applying the SFR laws in these regions provides a tool to consider the effect of the distance from the centre of the galaxy on SFR laws such as metallicity, gas mass, stellar mass gradients and changing the ratio of H$_2$ to H\,{\sc I}.

\subsection{Star formation laws}
\label{sec: sfl}

We made three SFR maps, three gas mass density maps and a stellar mass density map of the whole galaxy. We applied both laws on the whole galaxy using each combination of SFR and gas mass density tracer, along with applying the laws on three different regions. \citet{Kennicutt12} reviewed star formation in the Milky Way and nearby galaxies and described that there might be a breakdown of SFR laws for small size regions. Therefore, we re-gridded all the maps to have a pixel size equivalent to 750~pc, and repeated all of our processes on these new maps (Table~\ref{table:res750}). Additionally, we used these results to see if there is any correlation between the SFR laws and metallicity, as will be described in Section~\ref{sec:fittingmetal}.
 
\begin{figure*}

\includegraphics[width=168mm , height=230mm]{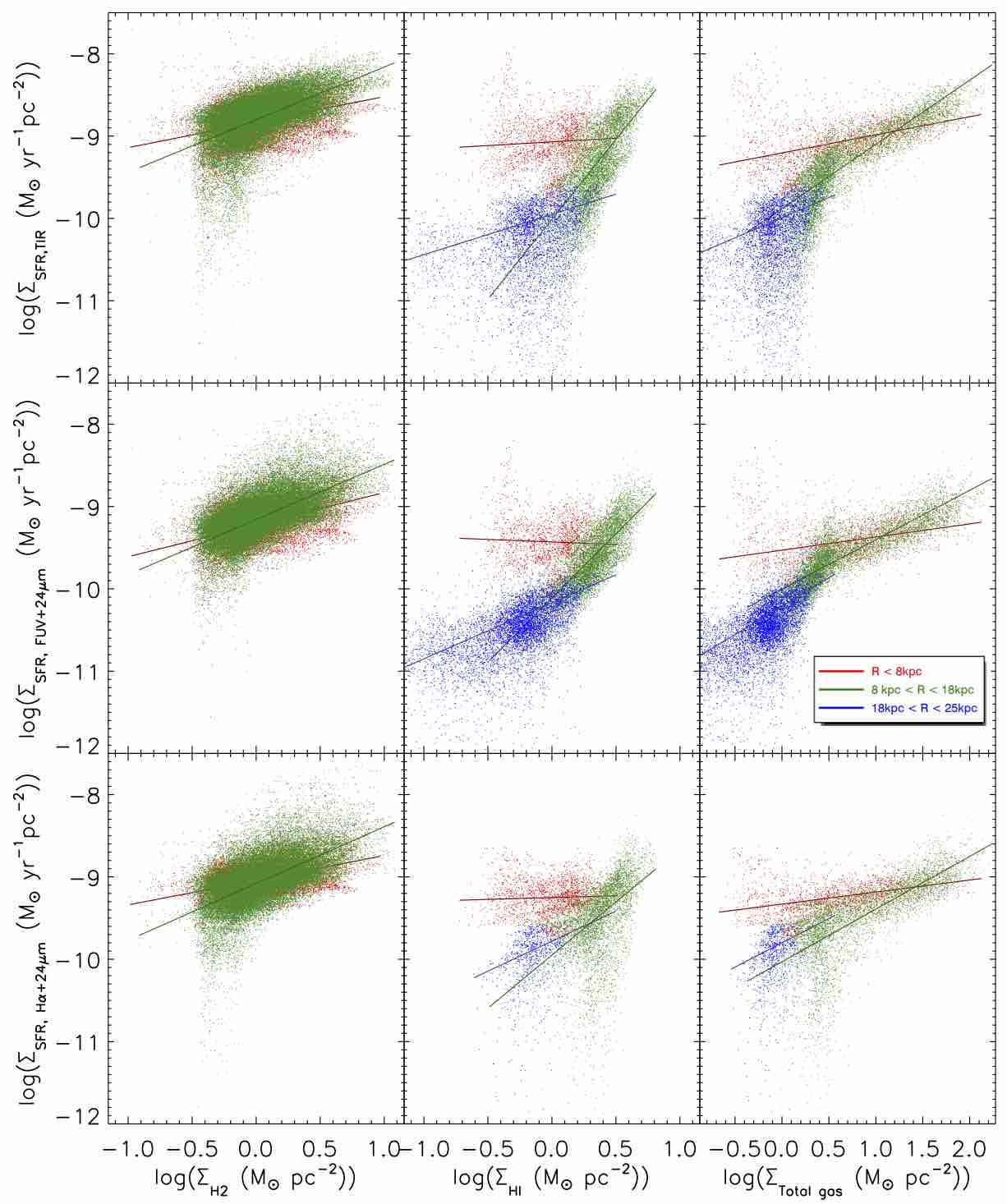}
\caption{Same as figure~\ref{fig:ks_all}, but in this figure we separated pixels from different regions in the galaxy by their colours. The regions with $R< 8\kpc$, $8\kpc < R < 18\kpc$, and $18\kpc < R \la 25\kpc$ are shown in red, green and blue, respectively.}
\label{fig:ks,regs}
\end{figure*}

Figure~\ref{fig:ks_all} shows the pixel-by-pixel fitting of the K-S law on the whole galaxy. The first row of plots are the surface density of the SFR(TIR), which is calculated using the TIR emission from Section~\ref{sec:sfr_fir}, versus the gas surface density traced by only the molecular gas (H$_2$), only atomic gas (H\,{\sc I}), and total gas from right to left, respectively. 
The only difference between the upper and the other rows is that in the lower rows the SFR indicator is FUV plus 24~$\mu$m (second row), and \halpha plus 24~$\mu$m emission (third row). Figure~\ref{fig:ks,regs} shows the same data as Figure~\ref{fig:ks_all}. However, in this figure we separated pixels from different regions in the galaxy by their colours. 
 There is a deviation in the middle panel plots of Figure~\ref{fig:ks_all} which shows a set of points with with high star formation rate and low H\,{\sc I} gas.
As can be clearly seen in middle panel of Figure~\ref{fig:ks,regs}, this deviation is due to regions within 8~kpc of the centre of the galaxy (red points).
 \cite{Braun09} saw a similar feature in their studies of the K-S law in M31. 
The H\,{\sc I} surface density profile of M31 \citep[see figure 16 in][]{Chemin09} shows several local minima from the centre of M31 out to 30~kpc distance which could be a result of the H\,{\sc I} wrap. The H\,{\sc I} surface density map shows a lower density in the central regions compared to the ring area; on the other hand, the SFR maps show high densities in the central regions. The combination of the relatively low gas and high SFR causes the deviation. However,~~\cite{Braun09} argued that the high 24~$\mu$m fluxes in the centre of the galaxy are due to other processes such as shock heating rather than a large amount of star formation.
Since the total gas mass is calculated by adding H\,{\sc I} and H$_2$ gas mass, we can see this deviation in the right panel of Figure~\ref{fig:ks_all} as well. However, since we have H$_2$ emission in those regions these deviations are less prominent than the ones in the middle panel.

We took a similar approach for the extended Schmidt law. Figures~\ref{fig:es,all,fuv,tot}-~\ref{fig:es,regs,fuv,tot} and Figures~\ref{fig:es,all,fir,h2} - ~\ref{fig:es,all,halpha,tot} show the results of the pixel-by-pixel fitting of the extended Schmidt law. For the extended Schmidt law, we plotted our results in 3D to illustrate the relationship between the surface density of the SFR, the gas mass surface density, and stellar mass density more clearly. In this series of plots, the $x$-axis is the gas mass surface density, either molecular gas (H$_2$), atomic gas (H\,{\sc I}), or total gas, the $y$-axis is the SFR(TIR), SFR(FUV + 24~$\mu$m) or SFR(H$\alpha$ + 24~$\mu$m), and the $z$-axis is the stellar mass surface density. The shadows of the data on each surface are also plotted, to have a more clear picture of correlations between components. The shadows on the $x-y$ surface are the same as in Figure~\ref{fig:ks_all}. 

\begin{figure*}
        \centering
        \includegraphics[width=\textwidth]{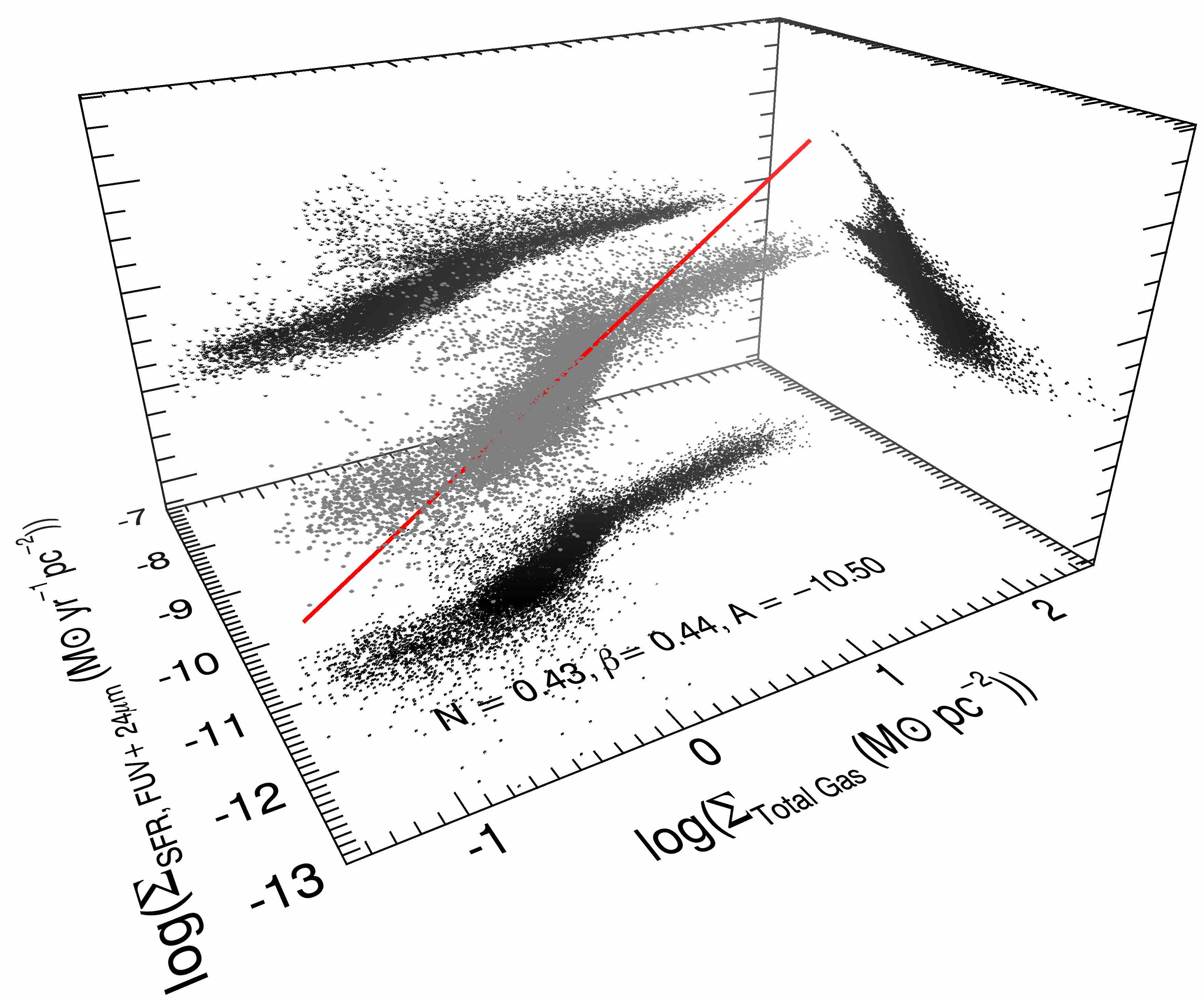}
        \caption{The result from fitting the extended Schmidt law on the surface density of SFR(FUV + 24~$\mu$m), the surface density of total gas and the surface density of stellar mass ($z$-axis) from the whole galaxy. Solid line shows the best fit (the mean value of the range reported in Table~\ref{table:res}) from a pixel-by-pixel method. Equivalent plots for the other SFR and gas tracers are in Figures~\ref{fig:es,all,fir,h2} - ~\ref{fig:es,all,halpha,tot} in the appendix.}
        \label{fig:es,all,fuv,tot}
    \end{figure*}

\begin{figure*}
\centering
\includegraphics[width=162mm]{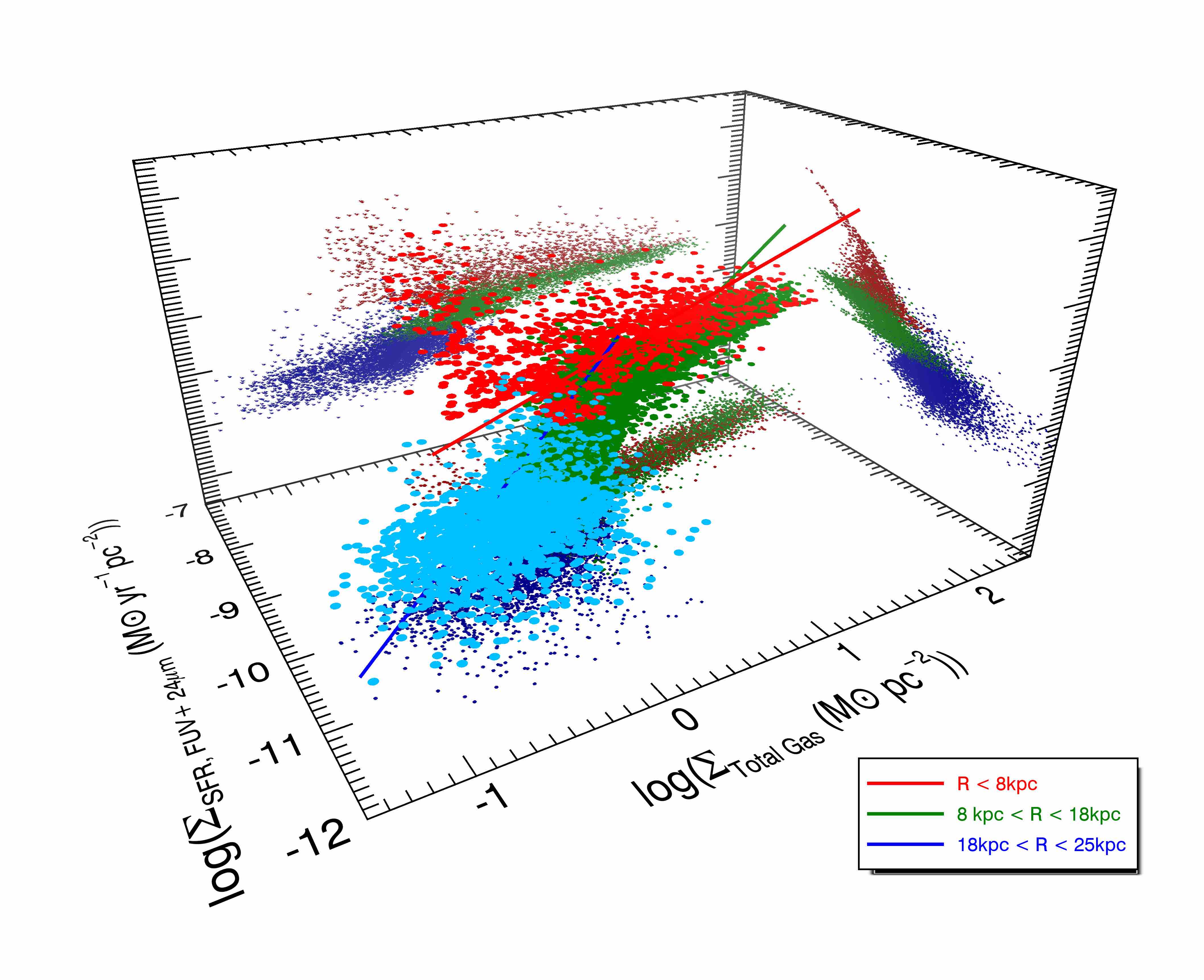}
\caption{Fitting result for SFR(FUV + 24~$\mu$m) vs total gas. Same as Figure~\ref{fig:es,all,fuv,tot}, but in this figure we denote pixels from different regions in the galaxy by their colours. The regions with $R< 8\kpc$, $8\kpc < R < 18\kpc$, and $18\kpc < R \la 25\kpc$ are shown in red, green and blue, respectively.}
\label{fig:es,regs,fuv,tot}
\end{figure*}

\begin{figure*}
\centering
\includegraphics[width=162mm]{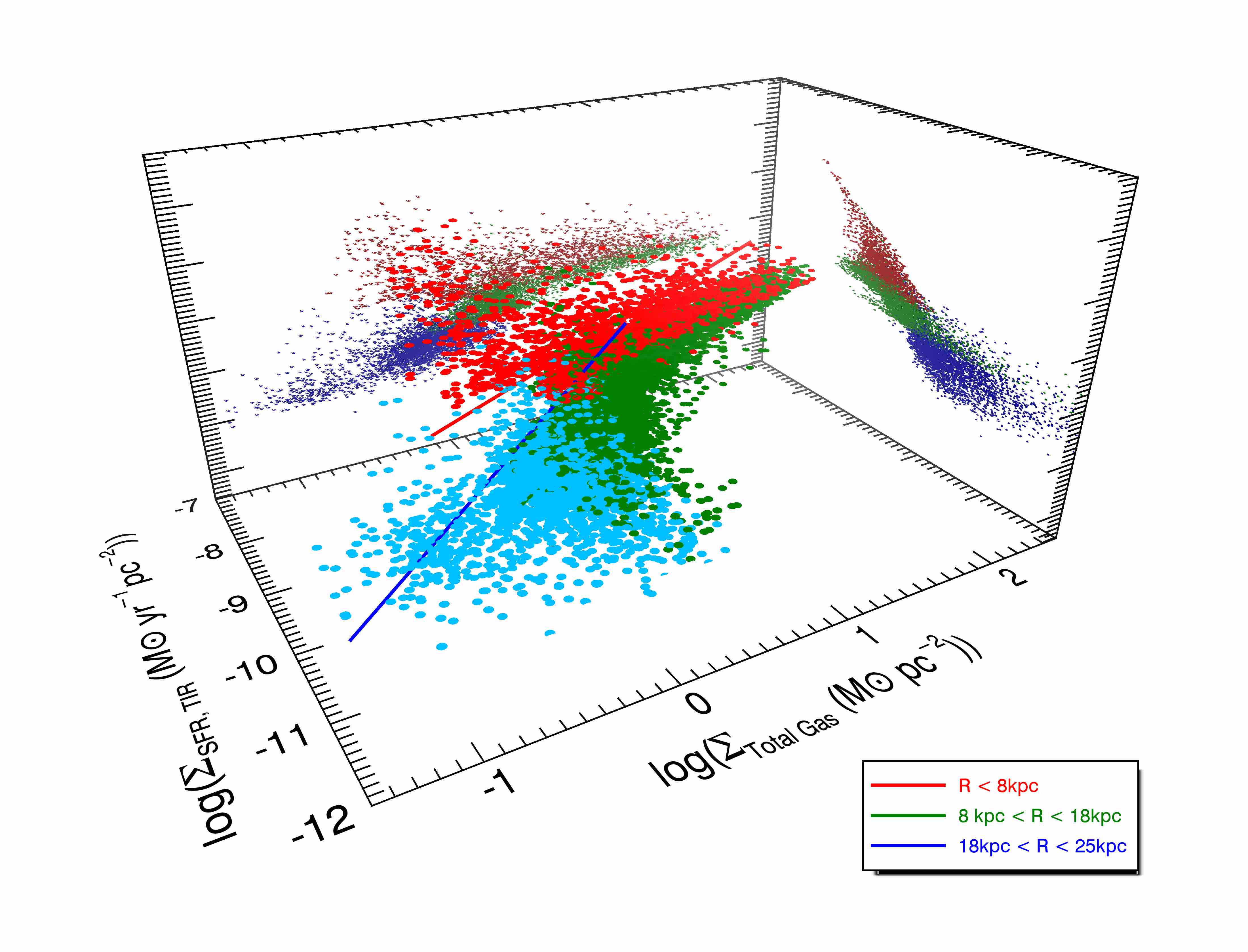}
\caption{Same as Figure~\ref{fig:es,regs,fuv,tot}, but in this figure we use total infrared emission as a tracer of the SFR.}
\label{fig:es,regs,fir,tot}
\end{figure*}

\begin{table*}
\center
\caption{ K-S law and Extended Schmidt law fitting results}
\label{table:res}
\begin{tabular}{ccccccccc}
\hline\hline
\multicolumn{1}{c}{\multirow{1}{*}{Region}} & SFR Tracer & Gas Tracer & {\it N}    & A      & \nprime & $\beta$ & A$^\prime$ \\
\hline\hline
\multicolumn{1}{c}{\multirow{9}{*}{Whole Galaxy}} &TIR & H$_2$ only & 0.57 - 0.58 & -8.88 - -8.87  & 0.51 - 0.53  &  0.31 - 0.33 & -9.42 - -9.39  \\
& TIR               & HI only    & 1.13 - 1.19 & -9.71 - -9.69   & 0.76 - 0.79 & 0.74 - 0.75 & -10.58 - -10.56 \\
& TIR               & Total gas  & 0.73 - 0.75 & -9.85 - -9.84   & 0.46 - 0.48 & 0.53 - 0.55 & -10.40 - -10.38 \\
& FUV + 24\um       & H$_2$ only & 0.61 - 0.63 & -9.24 - -9.23   & 0.57 - 0.59 & 0.20 - 0.22 & -9.60 - -9.57   \\
& FUV + 24\um       & HI only    & 1.16 - 1.19 & -9.97 - -9.95   & 0.78 - 0.81 & 0.59 - 0.60 & -10.64 - -10.62 \\
& FUV + 24\um       & Total gas  & 0.67 - 0.70 & -10.10 - -10.09 & 0.42 - 0.44 & 0.43 - 0.45 & -10.51 - -10.49 \\
& H$\alpha$ + 24\um & H$_2$ only & 0.60 - 0.62 & -9.14 - -9.13   & 0.54 - 0.56  & 0.33 - 0.35  & -9.72 - -9.70 \\
& H$\alpha$ + 24\um & HI only    & 0.46 - 0.54 & -9.64 - 9.61    & 0.48 - 0.53 & 0.72 - 0.76 & -10.75 - -10.69 \\
& H$\alpha$ + 24\um & Total gas  & 0.48 - 0.51 & -9.85 - 9.82    & 0.32 - 0.35 & 0.54 - 0.58 & -10.59 - 10.54    \\
\hline
\multicolumn{1}{c}{\multirow{9}{*}{R$< 8\kpc$}} & TIR & H$_2$ only & 0.30 - 0.33  & -8.84 - -8.83 & 0.31 - 0.33 & 0.67 - 0.70 & -10.28 - -10.23  \\
 & TIR               & HI only    & 0.02 - 0.16  & -9.09 - -9.06 & 0.74 - 0.83 & 0.84 - 0.89 & -10.83 - -10.73 \\
 & TIR               & Total gas  & 0.20 - 0.24  & -9.22 - -9.18 & 0.28 - 0.30 & 0.70 - 0.73 & -10.70 - -10.63 \\
 & FUV + 24\um       & H$_2$ only & 0.37 - 0.41  & -9.22 - -9.21 & 0.26 - 0.32 & 0.43 - 0.46 & -10.01 - -9.96  \\
 & FUV + 24\um       & HI only    & -0.13 - 0.01 & -9.44 - -9.42 & 0.67 - 0.76 & 0.81 - 0.86 & -10.64 - -10.62 \\
 & FUV + 24\um       & Total gas  & 0.13 - 0.18  & -9.55 - -9.51 & 0.24 - 0.26 & 0.68 - 0.73 & -11.03 - -10.94 \\
 & H$\alpha$ + 24\um & H$_2$ only & 0.29 - 0.32  & -9.14 - -9.13 & 0.38 - 0.41 & 0.50 - 0.53 & -10.33 - -10.26  \\
 & H$\alpha$ + 24\um & HI only    & 0.00 - 0.10  & -9.26 - 9.24  & 0.48 - 0.53 & 0.61 - 0.64 & -10.51 - -10.44 \\
 & H$\alpha$ + 24\um & Total gas  & 0.13 - 0.16  & -9.34 - 9.32  & 0.18 - 0.19 & 0.50 - 0.53 & -10.41 - -10.35     \\
\hline
\multicolumn{1}{c}{\multirow{9}{*}{$8\kpc < $R $< 18\kpc$}} & TIR & H$_2$ only & 0.63 - 0.65 & -8.80 - -8.79   & 0.47 - 0.49 & 0.68 - 0.70 & -9.97 - -9.93   \\
 & TIR               & HI only    & 1.90 - 1.99 & -10.03 - -10.00 & 1.33 - 1.39 & 0.92 - 0.96 & -11.08 - -11.03 \\
 & TIR               & Total gas  & 0.76 - 0.79 & -9.89 - -9.87   & 0.55 - 0.58 & 0.53 - 0.58 & -10.70 - -10.63 \\
 & FUV + 24\um       & H$_2$ only & 0.66 - 0.68 & -9.16 - -9.15   & 0.54 - 0.56 & 0.54 - 0.57 & -10.12 - -10.08 \\
 & FUV + 24\um       & HI only    & 1.52 - 1.59 & -10.12 - -10.09 & 1.12 - 1.17 & 0.60 - 0.63 & -10.64 - -10.62 \\
 & FUV + 24\um       & Total gas  & 0.60 - 0.61 & -10.01 - -10.00 & 0.48 - 0.50 & 0.29 - 0.34 & -10.36 - -10.31 \\
 & H$\alpha$ + 24\um & H$_2$ only & 0.69 - 0.70 & -9.09 - -9.08   & 0.53 - 0.55 & 0.70 - 0.73 & -10.30 - 10.25  \\
 & H$\alpha$ + 24\um & HI only    & 1.28 - 1.35 & -10.00 - -9.95  & 0.64 - 0.76 & 1.17 - 1.26 & -10.51 - -10.44 \\
 & H$\alpha$ + 24\um & Total gas  & 0.62 - 0.65 & -10.04 - -10.01 & 0.39 - 0.43 & 0.75 - 0.85 & -11.04 - -10.91  \\
\hline
\multicolumn{1}{c}{\multirow{9}{*}{$18\kpc <$ R $\la 25\kpc$}} & TIR & H$_2$ only &  N/A$^a$ & N/A$^a$ & N/A$^a$ &N/A$^a$ & N/A$^a$ \\
 & TIR               & H\,{\sc I} only    & 0.44 - 0.54 & -9.96 - -9.94  & 0.33 - 0.41    & 0.31 - 0.36    & -10.22 - -10.18     \\
 & TIR               & Total gas  & 0.44 - 0.54 &  -10.00 - -9.98  & 0.33 - 0.41  & 0.31 - 0.36    & -10.25 -10.21     \\
 & FUV + 24~$\mu$m       & H$_2$ only & N/A$^a$ & N/A$^a$ & N/A$^a$ &N/A$^a$ & N/A$^a$    \\
 & FUV + 24~$\mu$m       & H\,{\sc I} only    & 0.65 - 0.72 & -10.18 - -10.16  & 0.59 - 0.67    & 0.17 - 0.22    & -10.33 - -10.29     \\
 & FUV + 24~$\mu$m       & Total gas  & 0.64 - 0.72 & -10.22 - -10.21  &  0.59 - 0.67   & 0.16 - 0.22    & -10.37 - -10.33     \\
 & H$\alpha$ + 24~$\mu$m & H$_2$ only &  N/A$^a$ & N/A$^a$ & N/A$^a$ &N/A$^a$ & N/A$^a$     \\
 & H$\alpha$ + 24~$\mu$m & H\,{\sc I} only    & 0.48 - 0.97 &  -9.97 - -9.88  & 0.61 - 1.12    & -0.36 - -0.12    & -9.82 - -9.59     \\
 & H$\alpha$ + 24~$\mu$m & Total gas  & 0.40 - 0.90 & -10.01 - -9.95  &  0.53 - 1.03    & -0.35 - -0.11    & -9.82 - -9.59     \\
 \hline
\end{tabular}
\begin{tablenotes}
 \item $^a$  Since there are very few pixels which we can use for the fitting in this region we cannot provide a meaningful result.
\item Each entry shows the 95\% confidence range for the fitting parameter.
\item Fitting parameters of SF laws from applying the Bayesian method.  {\it N} is the power-law index of the K-S law; A is the intercept of the K-S law; \nprime is the gas power-law index in the extended Schmidt law;
 $\beta$ is the power-law index of the stellar component in the extended Schmidt law; and A$^\prime$ is the intercept of the extended Schmidt law.
\end{tablenotes}
\end{table*}

\begin{table*}
\center
\caption{K-S law and Extended Schmidt Law fits for larger pixels}
\label{table:res750}
\begin{tabular}{cccccccc}
\hline\hline
\multicolumn{1}{c}{\multirow{1}{*}{Region}}  & Gas Tracer & {\it N} & A  & \nprime & $\beta$ & A$^\prime$ \\
\hline\hline
\multicolumn{1}{c}{\multirow{3}{*}{Whole Galaxy}}
 & H$_2$ only & 0.26 - 0.31 & -10.00 - -9.91  & 0.21 - 0.26  & 0.26 - 0.40    & -10.51 - -10.30  \\
 & H\,{\sc I} only    & 1.08 - 1.24 & -11.60 - -11.36  & 0.73 - 0.85  & 0.58 - 0.64    & -11.74 - -11.60     \\
 & Total gas  & 0.66 - 0.73 & -10.11 - -10.06 & 0.39 - 0.46    & 0.41 - 0.49    & -11.56 - -10.47     \\
\hline
\multicolumn{1}{c}{\multirow{3}{*}{R$< 8\kpc$}}
 & H$_2$ only & 0.14 - 0.27 & -9.82 - -9.59  & 0.14 - 0.21   & 0.48 - 0.66   & -11.04 - -10.66      \\
 & H\,{\sc I} only    & 1.42 - 1.68 & -12.35 - -11.92 & 0.51 - 0.92    & 0.72 - 0.97    & -12.52 - -11.61     \\
 & Total gas  & 0.07 - 0.31 & -9.63 - -9.45  & 0.22 - 0.32    & 0.63 - 0.78    & -11.17 - -10.84      \\
\hline
\multicolumn{1}{c}{\multirow{3}{*}{$8\kpc < $R $< 18\kpc$}}
 & H$_2$ only & 0.27 - 0.32 & -9.98 - -9.86  & 0.17 - 0.24    &  0.38 - 0.73   & -10.89 - -10.44       \\
 & H\,{\sc I} only    & 0.88 - 1.38 & -11.80 - -11.06 & 1.00 - 1.22    & 0.53 - 0.69    & -12.41 - -12.09     \\
 & Total gas  & 0.55 - 0.64 & -10.03 - -9.96 & 0.40 - 0.51    & 0.27 - 0.49    & -10.53 - -10.29     \\
\hline
\multicolumn{1}{c}{\multirow{3}{*}{$18\kpc <$ R $\la 25\kpc$}} 
 & H$_2$ only & N/A$^a$& N/A$^a$ & N/A$^a$ &N/A$^a$ & N/A$^a$    \\
 & H\,{\sc I} only    & 0.46 - 0.81 & -11.21 - -10.80  & 0.44 - 0.76    & 0.13 - 0.36    & -11.35 -10.94     \\
 & Total gas  & 0.47 - 0.82 & -10.25 - -10.18 & 0.44 - 0.78    & 0.12 - 0.36    & -10.49 - -10.30     \\
 \hline
\end{tabular}
\begin{tablenotes}
\item Similar to Table~\ref{table:res} but here the fitting is performed on regions with a size of 750~pc, and we only show the results from SFR(FUV+24$\mu$m).
\end{tablenotes}
\end{table*}

%
\section{Discussion}

Table~\ref{table:res} shows the results of the hierarchical Bayesian fitting for the whole galaxy and all three regions. Considering the data and uncertainties for each parameter, we are 95\% confident that our fitting parameter is in the ranges given in Table~\ref{table:res}.
Our results suggest that changing the gas tracer has a more significant impact on the determination of the power law index than changing the SFR tracers. Therefore, we only discuss the results from SFR(FUV+24$\mu$m), and in Table~\ref{table:res750} we only show results from this SFR map.

\subsection{The K-S law in M31}

As mentioned before, the K-S law in M31 has been tested by many groups. In one of the most recent works on M31, \citet{Ford13} investigated the K-S law in six annuli and the global case using the ordinary least squares fitting method. 
To make a meaningful comparison between our results and the results reported in \citet{Ford13}, we averaged the power-law indices in \citet{Ford13} with a certain annuli and compared to the results with the same annuli in ours. We calculated power-law indices to be in the case of the H$_{2}$-only gas $N \sim 0.35, 0.58, 0.68$ and in case of total gas, $N \sim 1.37, 2.05, 1.60$ for $R< 8\kpc$, $8\kpc < R < 18\kpc$, and $18\kpc < R \la 25\kpc$, respectively.
\citet{Tabatabaei10} also applied the K-S law to M31 using the OLS fitting method, and have some differences with results from \citet{Ford13}. \citet{Ford13} argued that the main reason for these differences is the cut-off they used to fit the data. 
 
The whole-galaxy fit of the K-S law shows that using total gas or H$_{2}$-only gas gives a sub-linear relation between \sigmasfr and \sigmatotalgas. When using  H\,{\sc I}-only gas we find a nearly super-linear relation. These results are different from the power-law index estimated by \citet{Ford13}  in the case of total gas, $N=2.03\pm0.04$, but similar to their value of $N=0.6\pm0.01$ in the case of H$_{2}$-only gas. 
\citet{Tabatabaei10} estimated the K-S law power index as $N=1.30\pm0.05$ using total gas and $N=0.96\pm0.03$ using H$_{2}$-only gas. The reason behind these differences is mostly because of the statistical method we chose. The OLS fitting using H$_{2}$-only gas gives us a power-law index of $N=0.54\pm0.003$ which is close to the \citet{Ford13} results but still different fro \citet{Tabatabaei10}, perhaps for the same reason \citet{Ford13} suggested for the discrepancy between their results and results from \citet{Tabatabaei10}. We also noticed that different methods of measuring the uncertainties affect the power-law indexes, as well. The OLS fitting with \sigmatotalgas gives $N=0.74\pm0.003$, which is still lower than values in both  \citet{Tabatabaei10} and  \citet{Ford13}. Different methods in making the total gas map and using different criteria to choose outliers are the main reasons for this difference.

 \citet{Bigiel08} argued that there is no universal relationship between the surface density of the total gas and the SFR. They also showed that the SFR and the molecular hydrogen have a linear relationship. \citet{Shetty13} used the same data as \citet{Bigiel08} and argued that using the hierarchical Bayesian fitting leads to significant galaxy-by-galaxy variation between power-law indices with all of them being lower than those of \citet{Bigiel08}. Although some of our results in Table~\ref{table:res} showed a nearly linear correlation, we did not find any specific trend between power-law indices in the different regions.

The sub-linear power-law index in the case of the total gas or molecular gas surface density suggests that the star formation efficiency is decreasing as the total gas increases. Consequently, the depletion time increases. On the contrary, the super-linear relation suggests that the depletion time is decreasing with an increasing amount of H\,{\sc I} gas. The latter conclusion is more similar to the initial suggestion of the K-S law. If we assume that stars form in the molecular clouds, decreasing SFE with increasing total gas is justifiable.

The hierarchical Bayesian fitting on the $8<R<18$~kpc region gives similar values as the fitting to the whole galaxy data. This region contains the 10~kpc ring, which is where most of the star formation is happening, so it dominates the results from the whole galaxy.  For the other two regions, the K-S law fitting gives a sub-linear relation between the SFR and the gas surface density. A sub-linear relationship between \sigmasfr and  $\Sigma_{H_2}$ indicates that there are some CO-bright regions which do not associate with star forming regions.
 \citet{Krumholz09} showed that for regions with \sigmagas~$\leq 100$~M$_{\odot}$~pc$^{-2}$, \sigmasfr and $\Sigma_{H_2}$ have a nearly linear correlation. They also conclude that, at intermediate total gas column density, (\sigmagas~$\leq 100$~M$_{\odot}$~pc$^{-2}$), \sigmasfr and \sigmagas have a linear or slightly sub-linear correlation. The latter conclusion agrees with our results.

The K-S law power-law index varies with the distance to the centre of the galaxy. Similar to results from \citet{Ford13}, the power-law index considering molecular gas only increases with increasing distance from the centre of the galaxy. The power-law index using total gas or atomic gas does not show any particular correlation with distance. The increase of the power-law index while using molecular cloud only may be an effect of the decreasing metallicity across the galaxy or other physical properties of the galaxy such as stellar mass (see Section~\ref{sec:es_res}).
  
\subsection{The extended Schmidt law in M31}
\label{sec:es_res}
The extended Schmidt law has been a recently proposed law \citep{Shi11}, which shows a tight correlation between the SFR and the total gas surface density and the surface density of the stellar mass. Our results from fitting the whole galaxy shows a sub-linear relation between \sigmasfr and \sigmastar.
Using only atomic gas, $\beta$ is in good agreement with the original suggestion of the extended Schmidt law, $\eqnprime = 0.8 \pm 0.01$ and $\beta = 0.62\pm0.01$. However, $\beta$ is much smaller in other two gas tracers. Using total gas mass, we find $\eqnprime$ to be very close to $\beta$. This relation suggests that, although increasing the total gas causes a decrease to the SFE, the existing stars will increase the SFE. 

The fitting result on the $8<R<18$~kpc region (see Table~\ref{table:res}) is similar to that from fitting the whole galaxy together.  Taking the average $\beta$ from the atomic gas only fits gives us $\eqnprime \sim 0.79$ which is in good agreement with \citet{Shi11}, but $\beta = 0.53$  which is smaller than the original suggestion of the extended Schmidt law. However, the final indices reported by \citet{Shi11} for  the extended Schmidt law are an average over 12 different spiral galaxies, and some of their individual galaxies (e.g. NGC 4736, NGC 4726) have similar values as for M31.
Similar to the K-S law, the extended Schmidt law results are sensitive to the fitting method. Using the OLS fitting, one only will consider the SFR and stellar mass surface density errors, nevertheless, the hierarchical Bayesian regression fitting considers uncertainties on all the parameters.

Thus we can conclude that the SFR is higher in regions with higher stellar mass surface density than those with less stellar mass surface density.
In most cases this effect is almost as strong as the effect of gas surface density on the SFR.
\citet{Kim13}, using 3D numerical hydrodynamic simulations, showed that in the outer disk regions of a galaxy where the total gas is dominated by H\,{\sc I} gas, the SFR correlates with $\rho_{sd}^{0.5}$ where $\rho_{sd}$ is the mid-plane density of the stellar disk plus dark matter. This is because in the outer regions of the disk, the gravity from stars dominates. 
We calculated the power index related to the stellar mass surface density to be 0.36 for these regions (Table~\ref{table:res750}) which is slightly different from their suggestion. Since we do not have enough H$_2$ data for regions within $18\kpc < R \la 25\kpc$, our results are limited to ones from the extended Schmidt law for H\,{\sc I} gas only. Therefore, we could neither confirm nor refuse this prediction.

\subsection{Metallicity}
\label{sec:fittingmetal}
\begin{figure*}
\centering
\includegraphics[width=180mm]{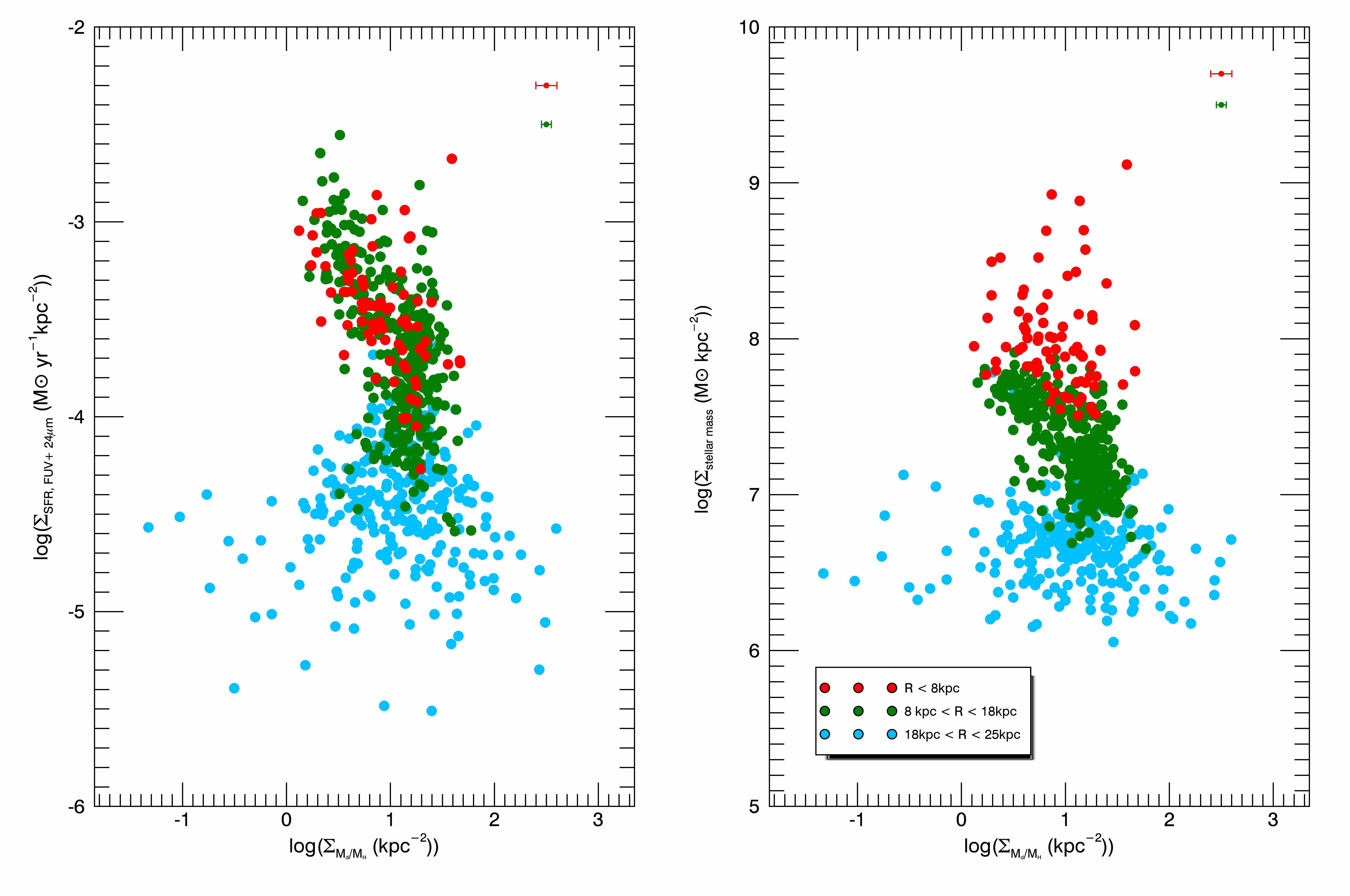}
    \caption{Left: the SFR(FUV + 24~$\mu$m) surface density versus gas/dust mass ratio surface density and right: Stellar mass versus gas/dust mass ratio surface density  shows the stellar mass. Each point shows a region with size of $\sim$750~pc. Typical uncertainties of the x-axes within 18~kpc are shown in upper right side of the plots. The regions with $R< 8\kpc$, $8\kpc < R < 18\kpc$, and $18\kpc < R \la 25\kpc$ are shown in red, green and blue, respectively.}
    \label{fig:metal}
\end{figure*}

It is well established that the metallicity of M31 is higher in the centre than it is in the outer disk  \citep[e.g.][]{Draine14}. The inverse correlation between metallicity and the X-factor, which was described in Section~\ref{sec:intro}, leads to an underestimation of the power-law indices from fitting the K-S law in the case of the molecular gas only and total gas. 
Thus a corrected power-law index of the K-S law is higher than shown in Table~\ref{table:res}. Another correlation between components of the SFR laws and metallicity was introduced by \citet{Krumholz09}. 
They found a new SFR law based on a correlation between SFR and metallicity, and suggested that this correlation strongly depends on the amount of total gas in galaxies. They also showed the SFR has a dependence on metallicity which we saw a weak correlation (Figure~\ref{fig:metal}).
 We estimated the SFR using \citet{Krumholz09} SFR laws for regions with size of $\sim$750~pc.
Using the mid value of metallicity, 1.65, and clumping factor of 1.2, We found that the SFR calculated using the Krumholz law is, on average, different by 45\% from the measured SFR(FUV + 24~$\mu$m) in this work. 

In more recent studies, \citet{Mannucci10} and \citet{Lilly13} assumed the K-S law holds for their sample of local and high-redshift galaxies and obtained a correlation between metallicity and the stellar mass, the gas mass and the SFR. 
They found that increasing both the SFR and the stellar mass increases metallicity, but increasing the gas mass decreases metallicity (equation 9 in \citep{Mannucci10}). 
However, by considering results from \citet{wong13} regarding the dependence of the rate of H\,{\sc I}-H$_2$ conversion to metallicity, \citet{Roychowdhury15} showed that the K-S law is independent of metallicity in galaxies with H\,{\sc I} as a dominant gas. 

In Figure~\ref{fig:metal}, we plot dust to gas mass ratio versus the SFR(FUV + 24~$\mu$m) and the stellar mass, and could not find any correlation over the scale of the whole galaxy.
We calculated the Pearson correlation coefficients between metallicity and SFR, and between metallicity and the stellar mass, and found them to be 0.08 and 0.04, respectively. 
However, in Figure~\ref{fig:metal}, we can see a correlation between metallicity and both SFR and stellar mass for regions within $R< 18\kpc$. These are weak correlations with Pearson correlation coefficients of $-0.58$ and $-0.46$ for the SFR case and the stellar mass case, respectively.
We cannot see any correlation in the data from $18\kpc < R \la 25\kpc$. The average metallicity uncertainty in this region is almost 10 times higher than the average uncertainty for the rest of the galaxy. This higher uncertainty obscures any tentative correlation between metallicity and SFR or stellar mass for from $18\kpc < R \la 25\kpc$.   
This analysis confirms that there is a correlation between metallicity and both the SFR/the stellar mass. Although we could not find any strong correlation between the SFR and metallicity, especially in outer disk regions, we should note that the effect of the metallicity on SF laws is ambiguous and more high resolution data is needed to solve this puzzle.

\section{Summary}
 We investigated the K-S law, the extended Schmidt law and the Krumholz law at both global and local scales in M31. We have determined the surface density of star formation rate, gas mass and stellar mass of this galaxy. Our three SFR maps in M31 use a combination of the FUV and 24~\um emission, a combination of the H$\alpha$ and 24~\um emission, and total infrared luminosity. We noticed that the H$_\alpha$ data available for M31 is not suitable for SFR studies (see appendix ~\ref{app:halpha}). We calculated the total SFR from FUV and 24 \um emission as 0.31$\pm$ 0.04 M$_{\odot}$yr$^{-1}$. We also produced the ISM map, using molecular gas only, atomic gas only and the total gas. Our main results are as follows:
\newline
1) The power-law index of the K-S law is mostly depended on which gas is used as a tracer of the gas mass in galaxy. Power-law indices are mostly independent of the SFR tracer.
\newline
2) Using different fitting methods gives different results. This dependence mostly comes from the way each fitting method handles the uncertainty on the star formation law parameters.
\newline
3) Although there is a correlation between \sigmasfr and \sigmagas, it is not the same in all regions in M31. The star formation laws predict more accurate results on regions with relatively higher star formation. 
\newline
4) We confirmed the suggestions of \citet{Shi11} that the surface density of stars has an impact on the SFR, and in regions with low gas surface brightness this impact is even more important. 
\newline
5) We performed statistical tests and found no correlation between the SFR, the stellar mass surface density and metallicity in the case of the whole galaxy. However there is a weak correlation between these quantities within 18~kpc of the centre of the galaxy.
\newline 
Obtaining images with higher resolution and maps with better coverage of this galaxy will enable continuation of these studies in finer detail.

The authors thank the referee for insightful comments which helped us to improve the manuscript.
The authors thank R. Shetty for giving us his hierarchical Bayesian fitting code, D. Kruijssen for his idea about 3D plots, L. Chemin for giving us the H\,{\sc I} map, K. Gordon for his latest version of the MIPS 24 \um and 70 \um maps, and M. Smith for access to the \Herschel data. We also want to thank E. Rosolowsky, B.T. Draine, P. Massey, K. Sandstrom, M. Azimlu, and G. Ford for their useful suggestions.
The authors acknowledge research support from the Natural Sciences and Engineering Research Council of Canada and from the Academic Development Fund of the University of Western Ontario. 
This research made use of Montage, funded by the National Aeronautics and Space Administration's Earth Science Technology Office, Computation Technologies Project, under Cooperative Agreement Number NCC5-626 between NASA and the California Institute of Technology. Montage is maintained by the NASA/IPAC Infra-red Science Archive.
\bibliographystyle{mnras}
\bibliography{ref}

\newpage
\appendix
\section{SFR from H$\alpha$ plus 24 \um: procedure and result}
\label{app:halpha}

As mentioned in Section~\ref{sec:data}, we used the H$\alpha$ data from the Nearby Galaxies Survey \citep{Massey07} to create one of our SFR maps. These data cover 10 overlapping fields across the disk of M31 in broad and narrow bands and were observed between August 2000 and September 2002; we obtained them  from the NOAO Science Archive. Making a spatially resolved SFR map of M31 using H$\alpha$ emission requires removal of background and stellar emission in each field, masking out foreground stars and saturated regions, creating a mosaic of H$\alpha$ continuum-subtracted images, and finally correcting for the flux contribution to the \halpha filter from the [N II] 6583~\AA\ line.

The Nearby Galaxies survey $R$-band images were used to remove the stellar emission from H$\alpha$ using the scaling factor between fluxes in these two band passes determined by \citet{Azimlu11}. They estimated continuum emission in each H$\alpha$ image, using photometric measurements in both \halpha and R-band images, and determined the scaling factor. These images were observed from a ground-based telescope over two years' time. Consequently, there is a non-negligible sky contribution in the background which must be estimated and removed from each image in both bands. Unfortunately, there is no nearby `blank-sky' image taken with the same instrument available from which to measure the sky contribution, which makes the removal of background emission even harder. The only remaining choice is to subtract the local background for each region. However, subtracting local backgrounds from each image will result in a non-smooth background in the final mosaic image.

\begin{figure*}
\centering
\includegraphics[width=164mm]{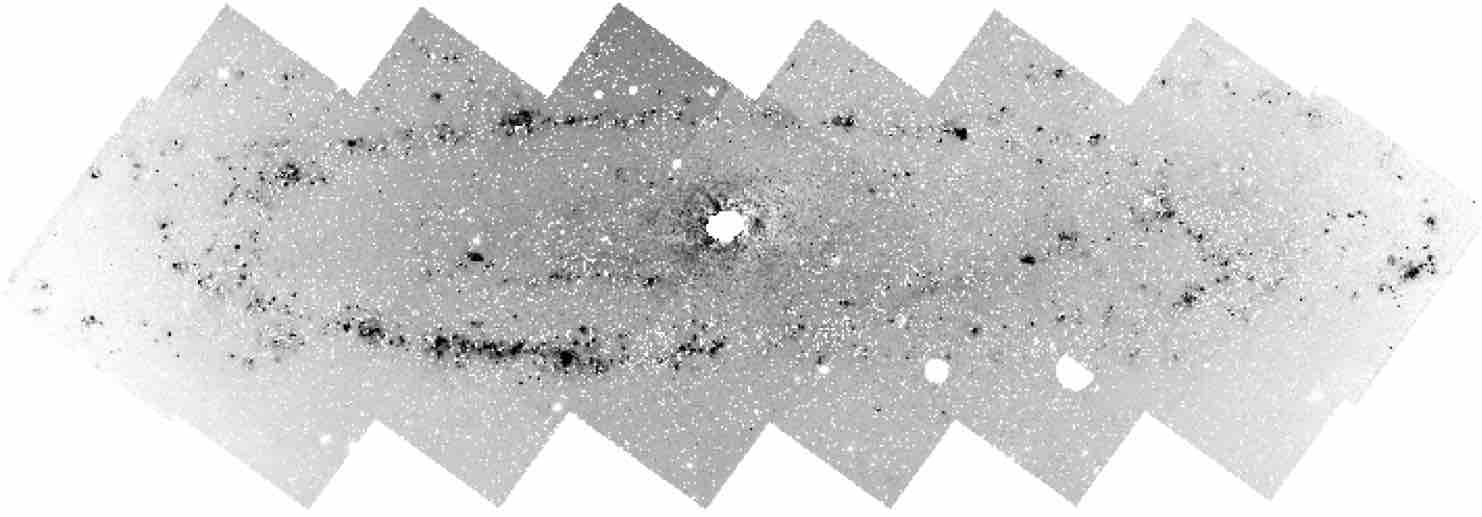}
\caption{Mosaic created using the Montage programme from six fields of H$\alpha$ emission images of M31 from \citet{Massey07}. The resulting image from Montage was continuum-subtracted and masked for all point sources. The centre of the galaxy was masked out due to saturation of data in the continuum $R$-band image.}
\label{fig:halpha}
\end{figure*}

To create the final mosaic, first we removed the background from each region in both the \halpha and $R$-band images. The second step was to subtract the continuum from the \halpha images. Since both \halpha and $R$-band images were aligned on the same coordinate grid, for each field, the R-band image multiplied by the scaling factor was subtracted from each corresponding \halpha image. At the end, we masked out all the foreground and saturated regions which include a $10\arcmin \times 10\arcmin$ region in the centre of the galaxy. To account for the flux contribution of the [N II] emission we used the flux ratio ${\rm [NII]}/{\rm H}\alpha = 0.54$ from \citet{Kennicutt08}, and subtracted it from the \halpha map. We created the final mosaic image, Figure~\ref{fig:halpha}, using the Montage program \citep{Berriman08}.

Creating the SFR map using \halpha plus 24 \um was described in Section~\ref{sec:sfr_halpha}. In order to investigate the SFR laws, we used the same method of the fitting as Section~\ref{sec:fitting}. The fitting results using SFR(\halpha $+$ 24~\um) are more or less the same as the other two SFR tracers. The main reason for this difference is the missing data in the centre of the galaxy and the lack of smooth background in the \halpha data. Thus we did not include the SFR(\halpha $+$ 24 \um) in our final analysis.

\newpage
\section{More results from the fitting of the extended Schmidt law}
\label{app:es,figs}
Section~\ref{sec: sfl} discusses testing the extended Schmidt law and shows results for one SFR/gas tracer combination. Here we present the plots for the remaining SFR/gas tracer combinations.

\begin{figure*}
    \centering
    \begin{subfigure}[b]{0.5\textwidth}
        \centering
        \includegraphics[width=\textwidth]{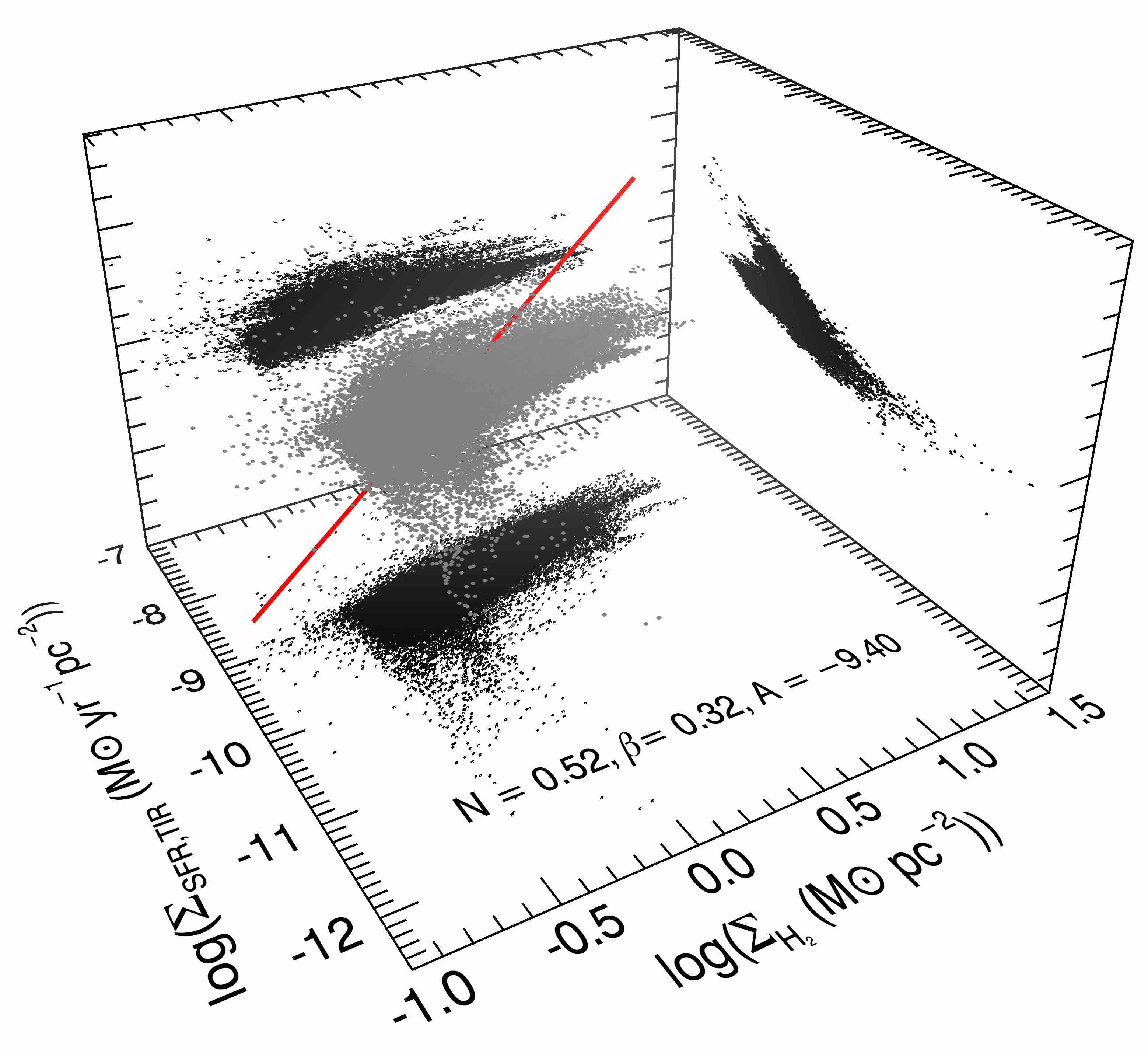}
        \caption{Surface density of SFR(TIR) vs surface density of H$_2$ and surface density of stellar mass ($z$-axis) }
        \label{fig:es,all,fir,h2}
    \end{subfigure}
    \hfill
    \begin{subfigure}[b]{0.5\textwidth}
        \centering
        \includegraphics[width=\textwidth]{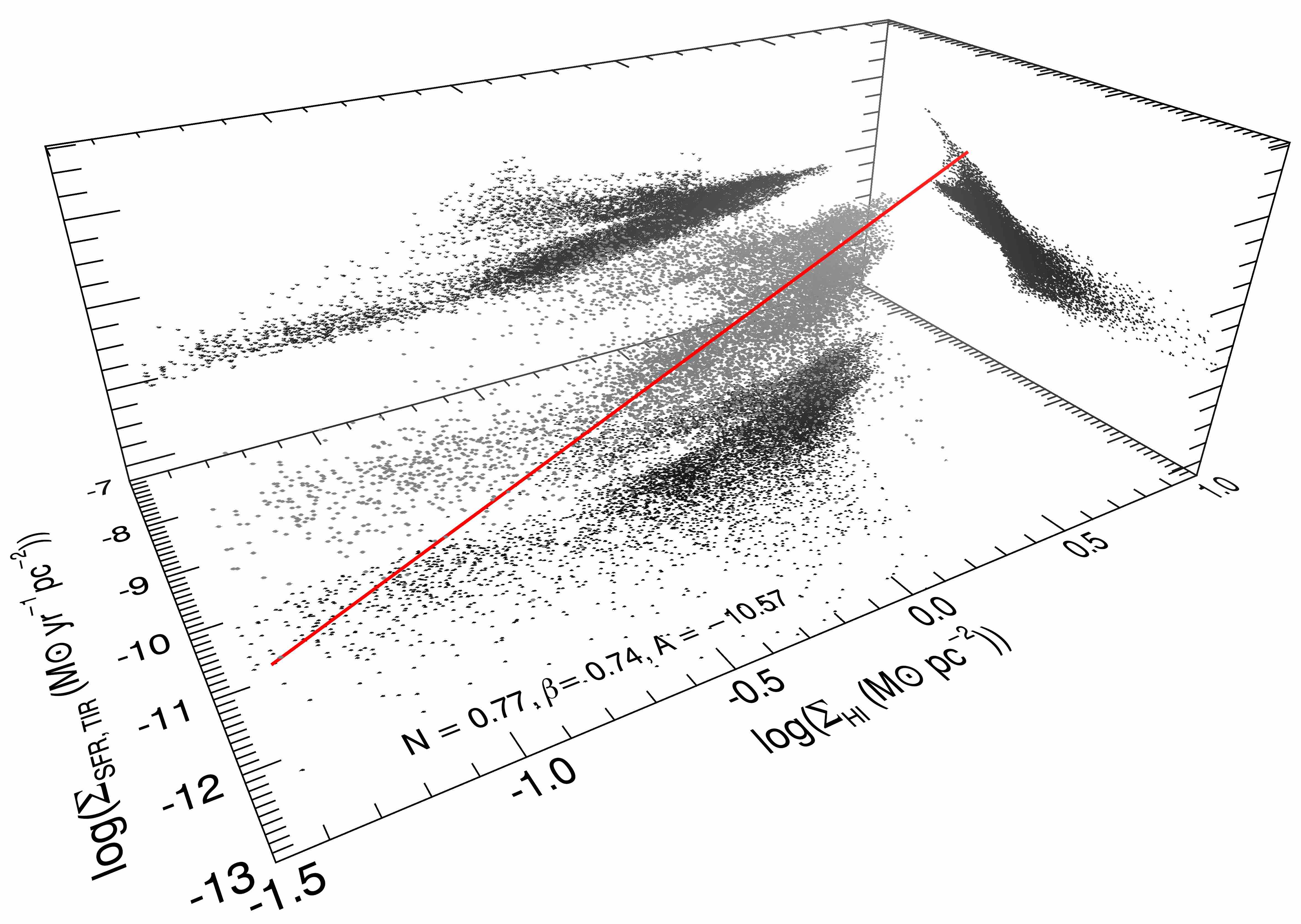}
        \caption{Surface density of SFR(TIR) vs surface density of H\,{\sc I} and surface density of stellar mass ($z$-axis) }
        \label{fig:es,all,fir,hi}
    \end{subfigure}
    \hfill
   \begin{subfigure}[b]{0.5\textwidth}
        \centering
        \includegraphics[width=\textwidth]{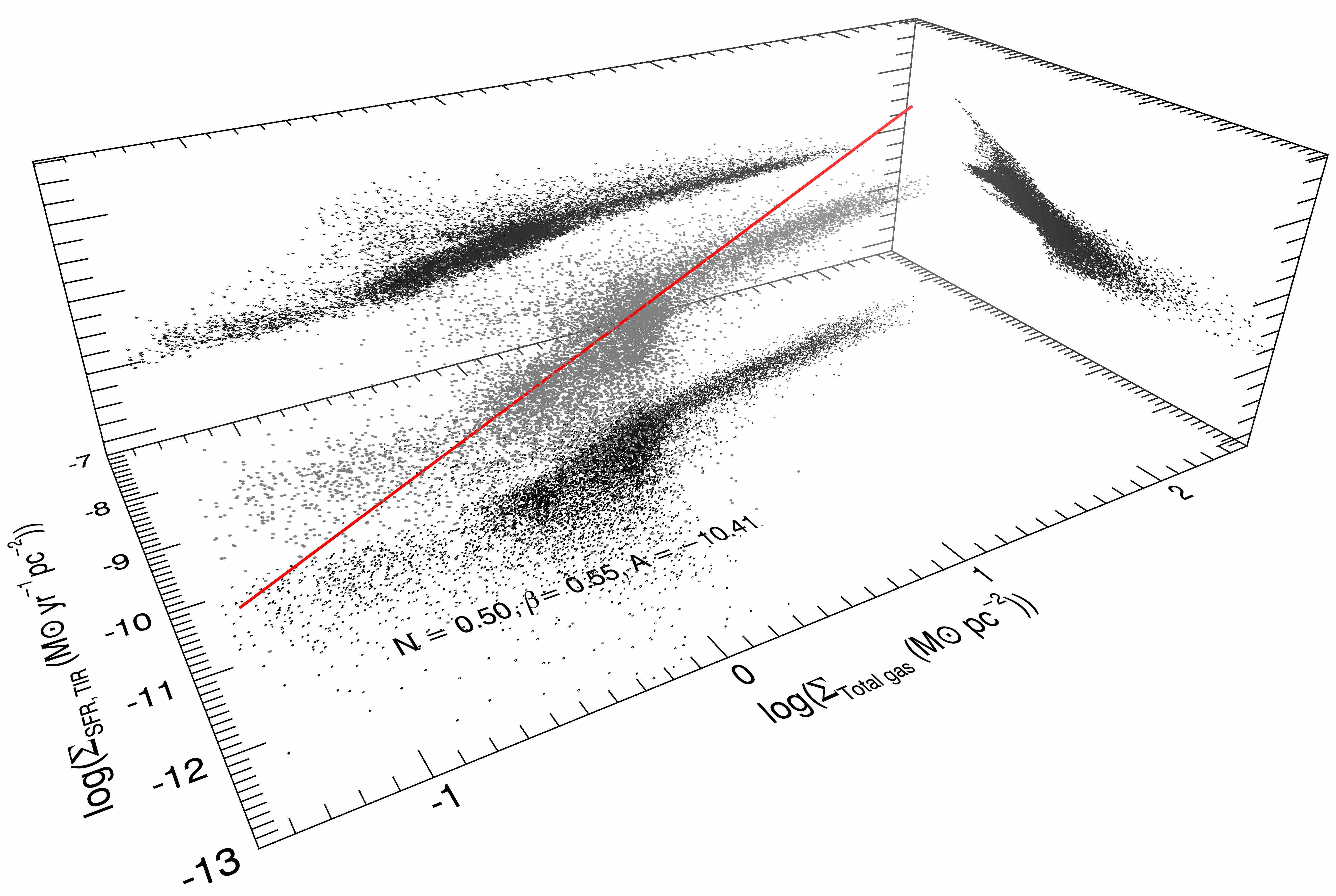}
        \caption{Surface density of SFR(TIR) vs surface density of total gas and surface density of stellar mass ($z$-axis)}
        \label{fig:es,all,fir,tot}
    \end{subfigure}
       \caption{Same as Figure~\ref{fig:es,all,fuv,tot}. Here, from top to bottom plots show the surface density of the SFR(TIR) vs. surface density of H$_2$, surface density of H\,{\sc I}, and the surface density of total gas, respectively. As in Figure~\ref{fig:ks_all}, the analyses use different pixel sizes: each point in the plots with the surface density of H$_2$ as a tracer of gas mass represents a region of size $\sim$30~pc and each point in the plots with the surface density of H\,{\sc I} or total gas mass represents a region of size $\sim$155~pc. Solid lines show the best fit, using the mean value of the ranges in Table~\ref{table:res}.}
       \label{fig:es,fir}
\end{figure*}

\begin{figure*}
    \centering
     \begin{subfigure}[b]{0.5\textwidth}
        \centering
        \includegraphics[width=\textwidth]{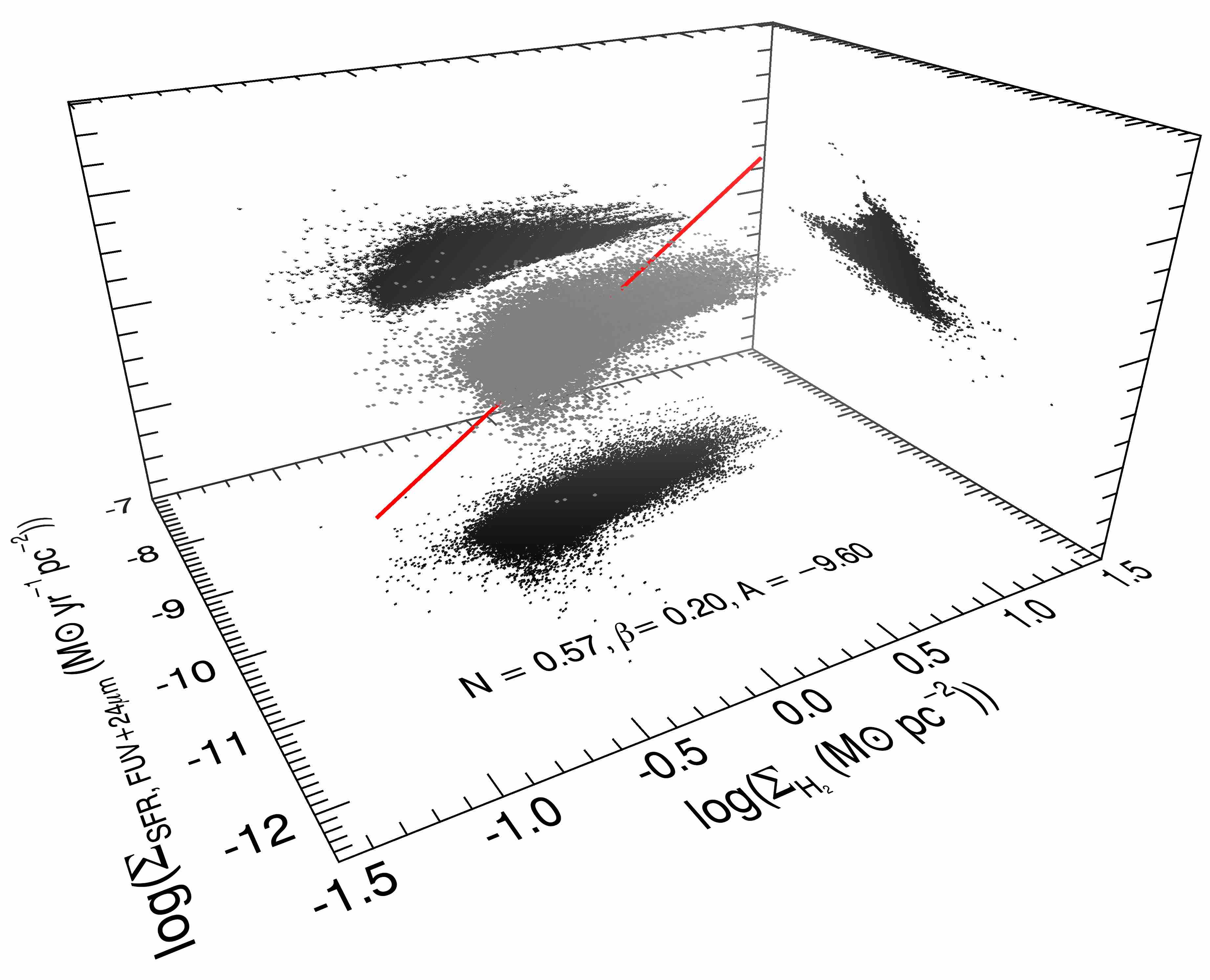}
        \caption{Surface density of SFR(FUV+24~$\mu$m) vs surface density of H$_2$ and surface density of stellar mass ($z$-axis)}
        \label{fig:es,all,fuv,h2}
    \end{subfigure}
     \hfill
   \begin{subfigure}[b]{0.5\textwidth}
        \centering
        \includegraphics[width=\textwidth]{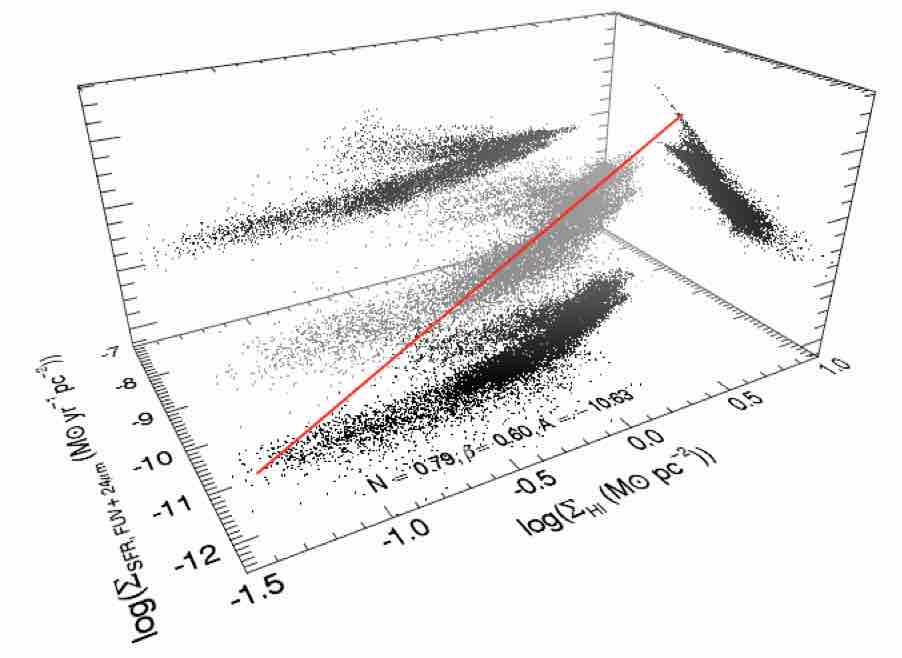}
        \caption{Surface density of SFR(FUV+24~$\mu$m) vs surface density of H\,{\sc I} and surface density of stellar mass ($z$-axis)}
        \label{fig:es,all,fuv,hi}
    \end{subfigure}
   \caption{Same as Figure~\ref{fig:es,fir}, but in this figure we used FUV + 24~$\mu$m as a tracer of the SFR. The plot of the surface density of the SFR(FUV+24~$\mu$m) vs the surface density of the total gas is shown in Figure~\ref{fig:es,all,fuv,tot} in the main text.}
\end{figure*}

\begin{figure*}
  \centering
   \begin{subfigure}[b]{0.5\textwidth}
        \centering
        \includegraphics[width=\textwidth]{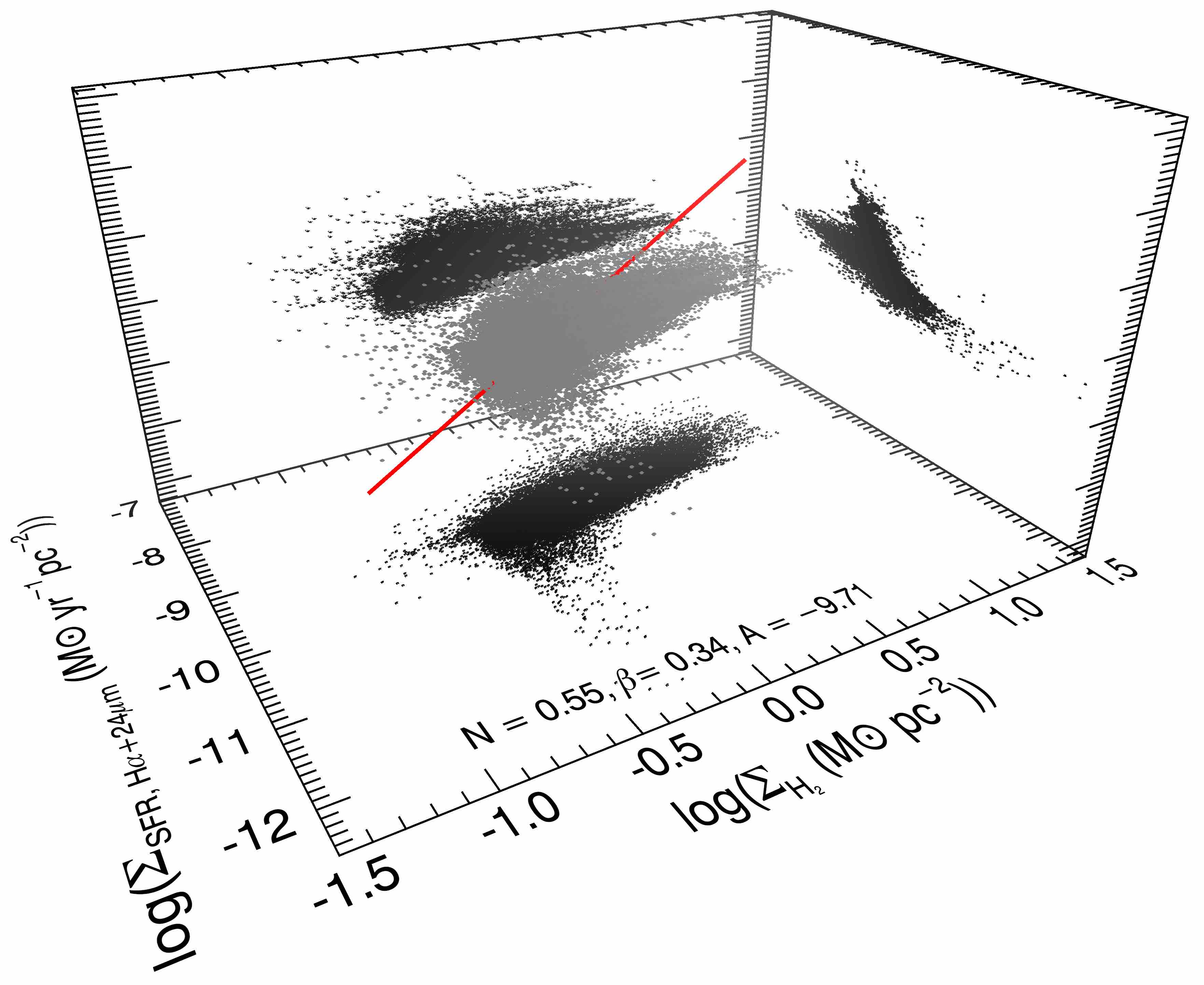}
        \caption{Surface density of SFR(H$\alpha$+24~$\mu$m) vs surface density of H$_2$ and surface density of stellar mass ($z$-axis)}
        \label{fig:es,all,halpha,h2}
    \end{subfigure}
     \hfill
      \begin{subfigure}[b]{0.5\textwidth}
        \centering
        \includegraphics[width=\textwidth]{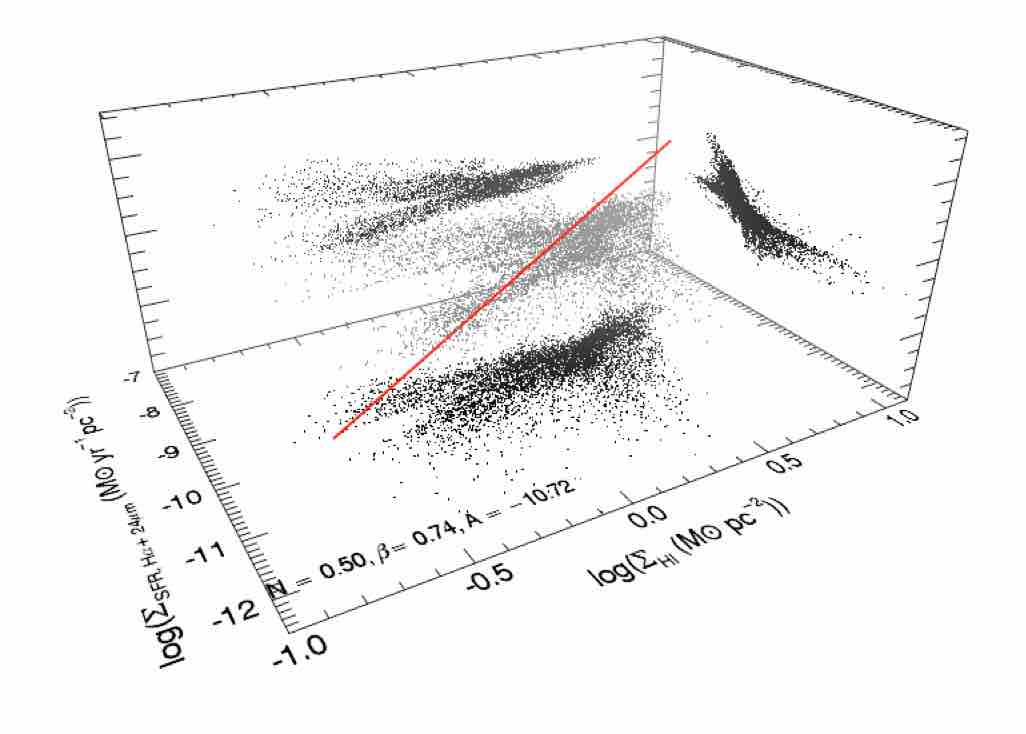}
        \caption{Surface density of SFR(H$\alpha$+24~$\mu$m) vs surface density of H\,{\sc I} and surface density of stellar mass ($z$-axis)}
        \label{fig:es,all,halpha,hi}
    \end{subfigure}
    \hfill
    \begin{subfigure}[b]{0.5\textwidth}
        \centering
        \includegraphics[width=\textwidth]{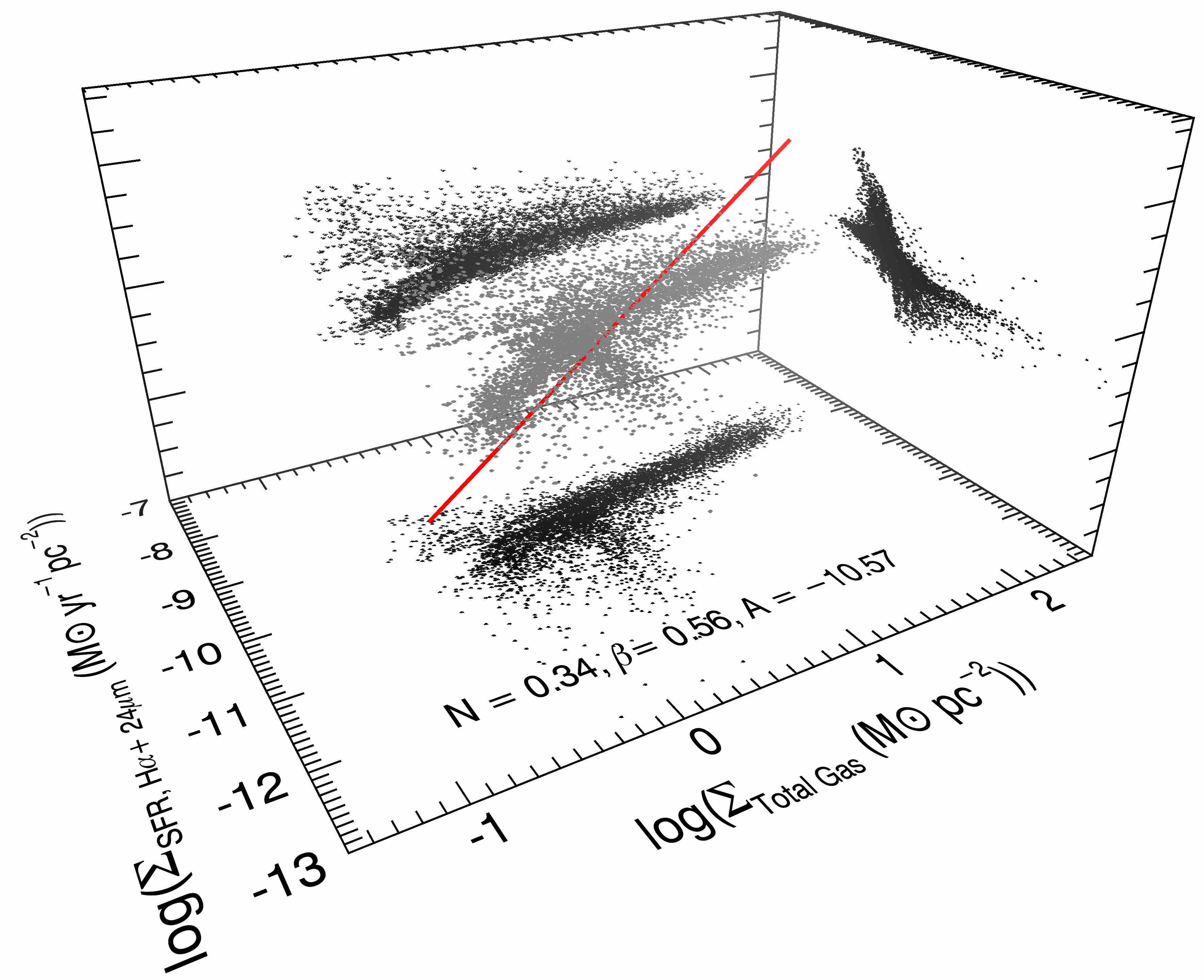}
        \caption{Surface density of SFR(H$\alpha$+24~$\mu$m) vs. surface density of total gas and surface density of stellar mass ($z$-axis)}
        \label{fig:es,all,halpha,tot}
    \end{subfigure}
    \caption{Same as Figure~\ref{fig:es,fir}, but in this figure we used H$\alpha$ + 24~$\mu$m as a tracer of the SFR.}
\end{figure*}

\end{document}